\documentclass[aps,prd,twocolumn,superscriptaddress,nofootinbib]{revtex4-1}

\usepackage{latexsym}
\usepackage{amsmath}
\usepackage{amssymb}
\usepackage{amsfonts}
\usepackage{bm}
\usepackage{physics}

\usepackage{color}
\definecolor{purple}{rgb}{0.5,0,0.5}
\definecolor{blue}{rgb}{0.0,0,0.9}
\definecolor{prdblue}{rgb}{0.133,0.118,0.498}
\usepackage[colorlinks=true, pdfstartview=FitV, linkcolor=prdblue, citecolor= prdblue, urlcolor=prdblue]{hyperref}

\usepackage{supertabular} 
\usepackage{placeins}
\usepackage{epsfig}
\usepackage{graphicx}

\begin{document}


\begin{flushright}
{\small JLAB-THY-23-3779}
\end{flushright}

\title{Light-front Wave Functions of Vector Mesons in an Algebraic Model}


\author{B. Almeida-Zamora}
\email[]{bilgai\_almeidaz@hotmail.com}
\affiliation{Departamento de Investigaci\'on en F\'isica, Universidad de Sonora, Boulevard Luis Encinas J. y Rosales, 83000, Hermosillo, Sonora, Mexico.}
\affiliation{Departamento de Sistemas F\'isicos, Qu\'imicos y Naturales, Universidad Pablo de Olavide, E-41013 Sevilla, Spain}

\author{J.J. Cobos-Mart\'inez}
\email[]{jesus.cobos@unison.mx}
\affiliation{Departamento de F\'isica, Universidad de Sonora, Boulevard Luis Encinas J. y Rosales, 83000, Hermosillo, Sonora, Mexico.}

\author{A. Bashir}
\email[]{adnan.bashir@umich.mx;abashir@jlab.org}
\affiliation{Instituto de F\'sica y Matem\'aticas, Universidad Michoacana de San Nicol\'as de Hidalgo, Morelia, Michoac\'an 58040, M\'exico.}
\affiliation{Theory Center, Jefferson Lab, Newport News, VA 23606, USA}

\author{K. Raya}
\email[]{khepani.raya@dci.uhu.es}
\affiliation{Departmento de Ciencias Integradas, Universidad de Huelva, E-21071 Huelva, Spain.}

\author{J. Rodr\'iguez-Quintero}
\email[]{jose.rodriguez@dfaie.uhu.es}
\affiliation{Departmento de Ciencias Integradas, Universidad de Huelva, E-21071 Huelva, Spain.}

\author{J. Segovia}
\email[]{jsegovia@upo.es}
\affiliation{Departamento de Sistemas F\'isicos, Qu\'imicos y Naturales, Universidad Pablo de Olavide, E-41013 Sevilla, Spain}


\date{\today}

\begin{abstract}
Inspired by the recent development of an algebraic model which provides an adequate and unified description of the internal structure of the lowest-lying pseudo-scalar mesons, belonging both to the light quarks sector and to the one of heavy quarks, we perform its first extension to the vector-meson case. The algebraic model describes meson's structure in terms of the spectral density function that appears in a Nakanishi integral representation of the covariant quark-antiquark bound-state amplitude, \emph{i.e.}, the Bethe-Salpeter amplitude. We compute the leading-twist light-front wave functions of the $\rho(770)$, $\phi(1020)$, $J/\psi$ and $\Upsilon(1S)$ mesons through their connection with the parton distribution amplitudes.
Among the results we present, the following are of particular interest: (i) transverse light-front wave functions can be obtained algebraically from the corresponding parton distribution amplitudes, whereas that is not the case for longitudinal light-front wave functions, which requires an intermediate step where a spectral density function must be derived from the particular parton distribution amplitude; (ii) the derived spectral density functions show marked differences between light and heavy vector mesons, the latter being narrower as compared to the former; these are also non-positive definite, although the integral over the entire curve is larger than zero as expected; and (iii) the longitudinal and transverse light-front wave functions of vector mesons with light quark content exhibit steep $x$- and $p_\perp^2$-dependence, while those of the $J/\psi$ and $\Upsilon(1S)$ mesons are characterized by narrow distributions in the $x$-range but, comparatively, much more gradual fall-offs with respect to the $p_\perp^2$-range depicted.
\end{abstract}


\maketitle


\section{Introduction}
\label{sec:intro}

Quantum Chromodynamics (QCD), the strongly interacting sector of the Standard Model of Particle Physics, presents a series of computational and conceptual challenges when its infrared solutions are sought. For example QCD's elementary excitations (quarks and gluons) are not those degrees-of-freedom directly accessible in experiment. These are instead confined inside color-singlet bound-states named hadrons. Another important fact of the theory is that there are numerous reasons to believe that QCD generates forces which are so strong that only less than 2\% of the nucleon's mass can be attributed to the so-called current-quark masses that appear in QCD's Lagrangian. The rest of its mass owes itself to the strong force capable of generating mass from literally nothing, an emergent phenomenon known as dynamical chiral symmetry breaking (DCSB). The properties of the ground-state pseudoscalar mesons, constituted from light quarks and simultaneously being the Goldstone bosons associated with the corresponding broken generators, are strongly influenced by DCSB; for instance, the quadratic increase of their masses is proportional to the masses of current quarks composing these mesons~\cite{Gell-Mann:1968hlm}.

Given that naive quantum mechanical models describe vector mesons as merely spin-flip excitations of their pseudo-scalar partners, it is natural to inquire into the impact of DCSB within the vector-meson bound states. One of the cleanest ways to expose differences between the impact of DCSB in pseudo-scalar and vector mesons is to compare their wave functions. However, such a comparison cannot adequately be investigated within non-relativistic quantum mechanics with a finite number of degrees of freedom; DCSB is an emergent phenomenon intimately related with the formulation of a quantum field theory. In this relativistic description, the object closest to a meson's wave function is the so-called Bethe-Salpeter wave function (BSWF)~\cite{Salpeter:1951sz}, which reduces to a Schr\"odinger wave function whenever a non-relativistic limit is sensible~\cite{Lucha:1998vh}; but that is never the case for the dressed valence-quarks of light mesons. One way to circumvent this problem is to compute the light-front wave function (LFWF) projection of the BSWF because the LFWF can translate features that arise purely through the infinitely many-body nature of relativistic quantum field theory into notions that are more familiar when formulating non-relativistic quantum mechanics~\cite{Brodsky:1997de}.

The BSWF can be obtained by a cumbersome computation that combines the Dyson-Schwinger equation (DSE) for the quark propagator and the Bethe-Salpeter equation (BSE) for mesons~\cite{Roberts:1994dr, Maris:1997tm, Chang:2009zb, Roberts:2011wy, Qin:2020jig}. This QCD based formalism has produced a plethora of theoretically interesting and experimentally testable quantities such as mesons masses~\cite{Chen:2012qr, Qin:2019oar}, their other static properties~\cite{Yin:2019bxe, Yin:2021uom}, form factors (FFs)~\cite{Chang:2013nia, Raya:2015gva} and parton distribution amplitudes (PDAs)~\cite{Gao:2014bca, Ding:2015rkn}. However, the calculation of, for instance, parton distribution functions (PDFs)~\cite{Ding:2019lwe, Cui:2020tdf}, generalized parton distributions (GPDs)~\cite{Mezrag:2014jka, Raya:2021zrz, Zhang:2021mtn} and transverse momentum distributions (TMDs)~\cite{Shi:2018zqd, Shi:2020pqe} remains a highly non-trivial task by employing full non-perturbvative QCD tools within the DSE-BSE formalism. Fortunately, our understanding of the intricate interplay between the DSE of the quark propagator and the meson's BSE~\cite{Binosi:2016rxz} allows us to construct models which are amicable enough to allow for algebraic manipulations and yet produce reliable predictions of the physical observables whose extraction from first principles remains a complicated problem.

The authors of Ref.~\cite{Albino:2022gzs} carried out the construction of an algebraic model for the quark propagator and the Bethe-Salpeter amplitude (BSA) of pseudo-scalar mesons in terms of a spectral density function (SDF). It leads to the derivation of the leading-twist LFWF by merely appealing to the definition of its Mellin moments. The resulting LFWF permits, on one hand, an algebraic connection with the PDA, so that the need to specify SDF is completely circumvented with prior knowledge of the PDA, and, on the other hand, the extraction of GPDs in the so-called DGLAP kinematic region through the overlap representation of the 
LFWFs~\cite{Diehl:2000xz}, and of a series of other distributions derived therefrom such as FFs and PDFs.

The proposed algebraic model adequately describes the internal structure of the lowest-lying pseudo-scalar mesons with either light or heavy quark content. That is to say, whenever a comparison is possible, the results showed agreement with other theoretical treatments such as lattice QCD, and also with experimental results. In this work, we extend such a study to investigate the internal structure of the $\rho(770)$, $\phi(1020)$, $J/\psi$ and $\Upsilon(1S)$ mesons. This exploratory work also allows us to identify the limitations of the algebraic model and the kind of physical observables that can be predicted in its present simple form.

This manuscript has been organized as follows: Section~\ref{sec:AlgMod} is devoted to the derivation of an algebraic model for the quark propagator and the BSA of vector mesons in terms of an SDF. Section~\ref{sec:PDAsLFWFs} contains a derivation of the relation between LFWFs and PDAs of vector mesons. There are two components of the LFWF, longitudinal and transverse; while the relation is straightforward in the latter case, one should connect the parallel components through the SDF. Section~\ref{sec:SDFs} describes how to obtain the SDF from a given valence-quark twist-two PDA of the vector meson. Our resulting LFWFs of vector mesons, from light to heavy quark sectors, are presented in Sec.~\ref{sec:results}; their most salient features are also discussed. Finally, we provide a brief summary, some concluding remarks, and potential prospects for future research in Sec.~\ref{sec:summary}.


\section{The Algebraic model}
\label{sec:AlgMod} 

Within a relativistic quantum field theory, the internal dynamics of a meson with spin-parity quantum numbers $J^P=1^-$ (vector meson) is described by its BSWF, $\chi_{v}$, which in turn is related to the associated BSA, $\Gamma_{v}$, and the quark (antiquark) propagator, $S_{q(\bar q')}$:
\begin{equation}
\chi_\mu^v(p,P) = S_q(p) \Gamma_\mu^v(p,P) S_{\bar q'}(p_{-}) \,,
\label{eq:BSWF}
\end{equation}
where $p_{-}=p-P$ and $P^2=-m_v^2$ with $m_v$ the mass of the vector meson. The labels $q$ and $\bar q'$ denote the valence quark and antiquark flavors, which can be different in general. Naturally, these flavors are the same for our analysis as it focuses on the $\rho(770)$, $\phi(1020)$, $J/\psi$ and $\Upsilon(1S)$ mesons, \emph{i.e.} the Lorentz-vector bound states with quark content $n\bar n$ ($n=u$ or $d$), $s\bar s$, $c\bar c$ and $b\bar b$, respectively.

Plain expressions for the quark (antiquark) propagator and BSAs that capture the essential features of our algebraic model are given by
\begin{align}
S_{q}(p) &= \Big(-i\gamma\cdot p + M_{q} \Big) \Delta(p^2, M_{q}^2) \,, \label{eq:Sq} \\
f_v \Gamma_\mu^v(p,P) &= i \gamma_{\mu}^{T} M_q \int_{-1}^{1} dw \; \rho_v (w) \Big[\hat{\Delta}(p_w^2,\Lambda_w^2) \Big]^\nu \label{eq:BSA} \,,
\end{align}
where $\Delta(s,t)\equiv (s+t)^{-1}$ and $\hat{\Delta}(s,t)\equiv t\Delta(s,t)$. Moreover, $\gamma_\mu^T=(\delta_{\mu\nu}-P_\mu P_\nu/P^2)\gamma_\nu$ is the transverse part with respect to momentum $P$ of the vertex, $f_v$ is the vector meson leptonic decay constant, $M_q$ is the dynamically dressed quark mass, the SDF is denoted by $\rho_v(w)$ and it defines the point-wise behavior of the BSA, $p_\omega=p+\frac{\omega}{2}P$ and
\begin{align}
\Lambda_w^2 = \Lambda^2(w) = M_q^2 - \frac{1}{4} (1 - w^2) m_v^2 \,.
\label{eq:Mw}
\end{align} 
As noticed in Ref.~\cite{Albino:2022gzs}, this particular functional $\omega$-dependence of $\Lambda$ leads to a simplification in relevant integrals and provides closed algebraic expressions which relate different structure distributions. Besides, Equation~\eqref{eq:Mw} has some additional conspicuous features that deserve to be highlighted: (i) a constant term, $M_q^2$, is retained, inherited from the kindred models~\cite{Raya:2022eqa, Raya:2021zrz, Zhang:2021mtn, Chavez:2021llq, Chavez:2021koz, Chouika:2017dhe, Chouika:2017rzs, Mezrag:2016hnp, Mezrag:2014jka, Xu:2018eii} that have been employed successfully to compute an array of GPD-related distributions; and (ii) all coefficients in the expression are chosen in such a way that the positivity of $\Lambda_\omega^2$ is guaranteed.

It is worth noting herein that the parameter $\nu>-1$ controls the asymptotic behavior of the BSA which must be ultraviolet finite since it resembles the wave function of a bound state~\cite{Maris:1999nt}; therefore, $\nu$ does not control any possible divergence. It just fits the asymptotic trend of the meson's BSWF. Moreover, $\nu=1$ is the most natural choice since it has been demonstrated that it yields the correct power law of the asymptotic behavior for mesons~\cite{Roberts:1994dr} and, in particular, $\nu=1$ recovers the results in Refs.~\cite{Chavez:2021koz, Chavez:2021llq, Chouika:2017rzs, Mezrag:2016hnp, Mezrag:2014jka}.

Inserting Eqs.~\eqref{eq:Sq} and~\eqref{eq:BSA} into~\eqref{eq:BSWF}, a straightforward algebraic manipulation leads us to a Nakanishi integral representation (NIR) of the BSWF:
\begin{equation}
f_v \chi_\mu^v(p,P) = M_q {\cal M}_{\mu;q,\bar q}^v(p_-,P) \int_{-1}^{1} dw \; {\cal D}_{q,\bar q}^\nu(p,P) \tilde{\rho}_v^\nu(w) \,,
\end{equation}
where the function ${\cal M}_{q,\bar q}^v(p_-,P)$ has the following tensor structure 
\begin{equation}
{\cal M}_{\mu;q,\bar q}^v(p_-,P) = (-i\gamma\cdot p + M_q)\gamma_{\mu}^T \left[-i\gamma\cdot p_- + M_{\bar q}\right] \,, 
\end{equation}
the function ${\cal D}_{q,\bar q}^\nu(p,P)$ is a product of quadratic
denominators
\begin{equation}
\mathcal{D}^\nu(q,P) = \Delta(q^2,M_q^2) \Delta(p^2_{w},\Lambda_\omega^2)^{\nu} \Delta(q_-^2,M_{\bar q}^2) \,,
\end{equation}
and the profile function $\tilde{\rho}_v^\nu(w)$ is defined in terms of the SDF as
\begin{equation}
\tilde{\rho}_v^\nu(w) \equiv \Lambda_w^{2\nu} \rho_v(w) \,.
\end{equation}
The three denominators in ${\cal D}_{q,\bar q}^\nu(p,P)$ can be combined into a single one using standard Feynman parametrization techniques,
\begin{align}
{\cal D}_{q,\bar q}^\nu(p,P) &= \nu (\nu+1) \int_0^1 d\beta \int_{\frac{1}{2}(\beta-1)(w-1)}^{\frac{1}{2}\left[(w+1)\beta-(w-1)\right]} d\alpha \nonumber \\
&
\times \frac{(1-\beta)^{\nu-1}}{\left[(p-\alpha P)^2 + \Omega_v^2\right]^{\nu+2}} \,,
\end{align}
here $\Omega_v^2=M_q^2 + \alpha(1-\alpha)P^2$ with $M_q=M_{\bar q}$. Finally, a suitable change of variables and a subsequent rearrangement in the order of integration yields the following expression for the BSWF:
\begin{align}
&
\chi_\mu^v(p,P) = \frac{\nu (\nu+1) M_q}{f_v} {\cal M}_{\mu;q,\bar q}^v(p_-,P) \nonumber \\
&
\times \int_{0}^{1} d\alpha \Bigg[ \int_{-1}^{1-2\alpha} dw \int_{\frac{2\alpha}{w-1}+1}^{1} d\beta + \int_{1-2\alpha}^1 dw \int_{\frac{2\alpha + w-1}{w+1}}^{1} d\beta \Bigg] \nonumber \\
&
\times \frac{(1-\beta)^{\nu-1}\tilde{\rho}_v^\nu (w)}{\left[(p-\alpha P)^2 + \Omega_v^2\right]^{\nu+2}} \,.
\label{eq:BSWF_final}
\end{align}
This expression serves as a starting point to compute the PDAs and LFWFs of vector mesons as detailed in the next section.


\section{Parton distribution amplitudes and light-front wave functions}
\label{sec:PDAsLFWFs}

A PDA is a LFWF of an interacting quantum system. For a meson, it can be written as $\phi(\xi;x)$ where $\xi$ is the momentum-scale that characterizes the exclusive process in which the meson is involved and $x$ expresses the light-front fraction of the bound-state total-momentum carried by the meson's valence quark, equivalent to the momentum fraction carried by the valence-quark in the infinite-momentum frame; momentum conservation entails that the valence antiquark carries the fraction $\bar x=(1-x)$. 

A vector meson has four leading-twist PDAs~\cite{Gao:2014bca}. The $\phi_\parallel(\xi;x)$ and $\phi_\perp(\xi;x)$ are usually named longitudinal and transverse PDAs, respectively, describing the light-front fraction of the meson's total momentum carried by the quark in a longitudinally or transversely polarized vector meson. The so-called $g_\perp^{(v)}(\xi;x)$ and $g_\perp^{(a)}(\xi;x)$ refer to transverse polarizations of quarks in the longitudinally polarized vector mesons. We would like to emphasize that only $\phi_\parallel(\xi;x)$ and $\phi_\perp(\xi;x)$ are independent at leading-twist since $g_\perp^{(v)}(\xi;x)$ and $g_\perp^{(a)}(\xi;x)$ are related to the longitudinal PDA~\cite{Ball:1998sk}:
\begin{subequations}
\begin{align}
\hspace*{-0.3cm} g_\bot^{(v)}(\xi;x) &=
\frac{1}{2}\left[\int^x_0 \! dv\frac{\phi_\|(\xi;v)}{\bar{v}}
+\int^1_x \!  dv\frac{\phi_\|(\xi;v)}{v}\right], \\
\hspace*{-0.3cm} g_\bot^{(a)}(\xi;x) &=
2 \left[\bar{x} \hspace{-1mm}  \int^x_0 \!  dv\frac{\phi_\|(\xi;v)}{\bar{v}} + x \hspace{-1mm} \int^1_x \!  dv\frac{\phi_\|(\xi;v)}{v} \right].
\end{align}
\end{subequations}
with $\bar{v}=1-v$ and $\bar{x}=1-x$. From now on, the scale-dependence of the PDAs, and of all other structure distributions, shall be taken into account only implicitly for the convenience of notation. The longitudinal and transverse PDAs of a vector meson are expressed as:
\begin{subequations}
\label{eq:PDAs}
\begin{align}
\phi_\parallel(x) &= \frac{N_\parallel m_v}{f_v} \nonumber \\
&
\times \Tr \int \frac{d^4p}{(2\pi)^4} \frac{\delta_n^x(p)}{n\cdot P} (\gamma\cdot n) n_\mu \chi_\mu^v (p,P) \,, \label{pdaparallel0} \\
\phi_\perp(x) &= \frac{N_\perp}{f_v^\perp m_v^2} \nonumber \\
&
\times \Tr \int \frac{d^4p}{(2\pi)^4} (n\cdot P) \delta_n^x(p) \sigma_{\mu \nu} P_\mu \chi_\nu^v (p,P) \,, \label{pdaperp0}
\end{align}
\end{subequations}
where $N_\parallel$ and $N_\perp$ are normalization constants; $f_v$ and $f_v^\perp$ are the meson's vector and tensor decay constants, with the former, a renormalization point invariant, explaining the strength of the vector meson decaying into an electron-positron pair, while the tensor decay constant depends
on the mass scale relevant to the process in which the meson is involved. The trace is taken over color and spinor indices, $\delta_n^x(p)=\delta(n  \cdot p - x \, n \cdot  P)$ with $n$ a light-like four-vector, such that $n^2=0$ and $n\cdot P=-m_{p}$. Finally, throughout this manuscript, dimensionless and unit normalized PDAs, $\int_0^1 dx\; \phi(x)=1$, shall be considered.

With the BSWF at hand, it is straightforward to follow the procedure explained originally in Ref.~\cite{Chang:2013pq} and thereby obtain the longitudinal and transverse PDAs from Eqs.~\eqref{eq:PDAs}. The first step is to compute the Mellin moments, defined as
\begin{equation}
\expval{x^m}_\phi = \int_0^1 dx \; x^m \, \phi (x) \,.
\end{equation}
For the parallel PDA, these are given by
\begin{eqnarray}
&& \hspace{-1cm} \expval{x^m}_{\phi_\parallel} = \frac{N_\parallel m_v}{f_v} \nonumber \\
&& \times \Tr \int \frac{d^4p}{(2\pi)^4} \frac{(n \cdot p)^m}{(n\cdot P)^{m+2}} (\gamma\cdot n) n_\mu \chi_{\mu}^{v}(p,P) \,.
\end{eqnarray}
Inserting Eq.~\eqref{eq:BSWF_final} into this expression, one arrives at:
\begin{align}
&
\expval{x^m}_{\phi_\parallel} = \frac{N_\parallel \nu(\nu+1) m_v M_q}{f_v^2} \nonumber \\
&
\times \Tr \int \frac{d^4p}{(2\pi)^4} \frac{(n \cdot p)^m}{(n\cdot P)^{m+2}} \gamma \cdot n n_\mu {\cal M}_{\mu;q,\bar q}^v(p,P) \nonumber \\
&
\times \int_{0}^{1} d\alpha \; \Bigg[ \int_{-1}^{1-2\alpha}\mathrm{d}w \int_{\frac{2\alpha}{w-1}+1}^{1} \mathrm{d}\beta \nonumber \\
&
+ \int_{1-2\alpha}^1\mathrm{d}w \int_{\frac{2\alpha + w-1}{w+1}}^{1} \mathrm{d}\beta \Bigg] \frac{(1-\beta)^{\nu-1}\tilde{\rho}_v^\nu (w)}{\left[(p-\alpha P)^2 + \Omega_v^2\right]^{\nu+2}} \,.
\end{align}
After the color and Dirac traces, the Mellin moments become
\begin{align}
&
\expval{x^m}_{\phi_\parallel} = \frac{4 N_c N_\parallel \nu(\nu+1) m_v M_q}{f_v^2}  \int_{0}^{1} d\alpha \nonumber \\
&
\times \left[ \int_{-1}^{1-2\alpha} dw \int_{\frac{2\alpha}{w-1}+1}^{1} d\beta + \int_{1-2\alpha}^1 dw \int_{\frac{2\alpha + w-1}{w+1}}^{1} d\beta \right] \nonumber \\
&
\times (1-\beta)^{\nu-1}\tilde{\rho}_v^\nu (w) \int \frac{d^4p}{(2\pi)^4} \frac{(n \cdot p)^m}{(n\cdot P)^{m+2}} \nonumber \\
&
\times \frac{M_q^2 + \left(p - \frac{n\cdot p}{n\cdot P}P\right)^2 - (n\cdot p)^2+ (n\cdot P)(n\cdot p)}{\left[(p-\alpha P)^2 + \Omega_v^2\right]^{\nu+2}} \,.
\end{align}
One may conveniently decompose the momentum integral into two parts, parallel and perpendicular:
\begin{equation}
\int \frac{d^4p}{(2\pi)^4} = \int \frac{d^2p_\perp}{16\pi^3} \int \frac{d^2p_\parallel}{\pi} \,,
\end{equation}
in such a way that the Mellin moments for the LFWF, expressed as
\begin{equation}
\label{eq:MM-LFWF}
\expval{x^m}_{\psi} = \int_0^1 dx\; x^m \, \psi(x,p_\perp^2=\text{fixed}) \,,
\end{equation}
can be obtained as follows:
\begin{align}
&
\expval{x^m}_{\psi_\parallel} = \frac{ 4N_c N_\parallel \nu(\nu+1) m_v M_q}{f_v^2}  \int_{0}^{1} d\alpha \nonumber \\
&
\times \left[ \int_{-1}^{1-2\alpha}dw \int_{\frac{2\alpha}{w-1}+1}^{1} d\beta + \int_{1-2\alpha}^1 dw \int_{\frac{2\alpha + w-1}{w+1}}^{1} d\beta \right] \nonumber \\
&
\times (1-\beta)^{\nu-1}\tilde{\rho}_v^\nu (w) \int \frac{d^2p_\parallel}{\pi} \frac{(n \cdot p)^m}{(n\cdot P)^{m+2}} \nonumber \\
&
\times \frac{M_q^2 + \left(p - \frac{n\cdot p}{n\cdot P}P\right)^2 - (n\cdot p)^2 + (n\cdot P)(n\cdot p)}{\left[(p-\alpha P)^2 + \Omega_v^2\right]^{\nu+2}} \,.
\end{align}
Performing the change of variable $p\to p + \alpha P$, decomposing the integrated momentum into its parallel and perpendicular parts, $p=p_\parallel + p_\perp$, such that
\begin{subequations}
\begin{align}
&
n \cdot p = n \cdot p_\parallel \,, \\
&
n \cdot p_\perp = 0 \,, \\
&
p^2 = p_\parallel^2 + p_\perp^2 \,,
\end{align}
\end{subequations}
and carrying out the integral over $p_\parallel$, the Mellin moments can be cast in the following form
\begin{align}
&
\expval{x^m}_{\psi_\parallel} = \frac{2N_c N_\parallel \nu M_q}{f_v^2 m_v}  \int_{0}^{1} d\alpha \alpha^m \nonumber \\
&
\times \left[ \int_{-1}^{1-2\alpha} dw \int_{\frac{2\alpha}{w-1}+1}^{1} d\beta + \int_{1-2\alpha}^1 dw \int_{\frac{2\alpha + w-1}{w+1}}^{1} d\beta \right] \nonumber \\
&
\times (1-\beta)^{\nu-1} \tilde{\rho}_v^\nu(w) \Bigg[ \frac{M_q^2 + p_\perp^2 - \alpha(1-\alpha) m_v^2}{\left( p_\perp^2 + \Omega_v^2 \right)^{\nu+1}} \nonumber \\
&
+ \frac{1}{2\nu} \frac{1}{\left( p_\perp^2 + \Omega_v^2 \right)^{\nu}} \Bigg] \,.
\end{align}
Finally, from the definition of the Mellin moments of the LFWF, Eq.~\eqref{eq:MM-LFWF}, the parallel LFWF of a vector meson is
\begin{align}
\label{eq:LFWF_para}
&
\psi_\parallel(x,p_\perp^2) = \frac{2 N_c N_\parallel \nu M_q}{m_{v} f_v^2} \times \nonumber \\ 
&
\times \left[ \int_{-1}^{1-2x} dw \int_{\frac{2x}{w-1}+1}^{1} d\beta + \int_{1-2x}^1 dw \int_{\frac{2x + w-1}{w+1}}^{1} d\beta \right] \nonumber \\
&
\times (1-\beta)^{\nu-1} \tilde{\rho}_v^\nu(w) \Bigg[ \frac{M_q^2 + p_\perp^2 + x(1-x) m_v^2}{\left( p_\perp^2 + \Omega_v^2 \right)^{\nu+1}} \nonumber \\
&
+ \frac{1}{2\nu} \frac{1}{\left( p_\perp^2 + \Omega_v^2 \right)^{\nu}} \Bigg] \,,
\end{align}
where $\Omega_v^2$ is now a function of $x$ instead of $\alpha$. Integrating out the $p_\perp$ dependence of $\psi_\parallel(x,p_\perp^2)$ yields the vector-meson's parallel PDA,
\begin{equation}
\phi_\parallel(x) = \int \frac{d^2p_\perp}{16\pi^3} \; \psi_\parallel (x,p_\perp^2) \,.
\end{equation}
Performing the shift $\nu \rightarrow 1+\delta$ to remove the spurious logarithmic divergence in $p_\perp$,\footnote{See text at the end of this section and the Appendix for further details about the way of dealing with this issue and its consequences on the final results.} it acquires the following form:
\begin{align}
\label{eq:PDA_para}
\phi_\parallel(x) &= \frac{N_c N_\parallel M_q}{8 \pi^2 m_v f_v^2} \nonumber \\
&
\times \left[ \int_{-1}^{1-2x} dw \int_{\frac{2x}{w-1}+1}^{1} d\beta + \int_{1-2x}^1 dw \int_{\frac{2x + w-1}{w+1}}^{1} d\beta \right] \nonumber \\
&
\times (1-\beta)^{\delta}\tilde{\rho}_v^{\delta +1}(w) \left[\frac{M_q^2 + x(1-x) m_v^2}{(\Omega_v^2)^{\delta+1}} + \frac{3}{2\delta} \frac{1}{(\Omega_v^2)^{\delta}} \right] \,.
\end{align}

A similar calculation follows for the transverse LFWF and PDA of a vector meson, leading to the following expressions, respectively,
\begin{align}
\label{eq:LFWF_perp}
&
\psi_\perp(x,p_\perp^2) = \frac{6 N_c N_\perp \nu M_q^2}{f_v f_v^\perp} \nonumber \\
&
\times \left[ \int_{-1}^{1-2x} dw \int_{\frac{2x}{w-1}+1}^{1} d\beta+ \int_{1-2x}^1 dw \int_{\frac{2x + w-1}{w+1}}^{1} d\beta \right] \nonumber \\
&
\times \frac{(1-\beta)^{\nu-1} \tilde{\rho}_v^\nu(w)}{\left(p_\perp^2 + \Omega_v^2\right)^{\nu+1}} \,,
\end{align}
and
\begin{align}
\label{eq:PDA_perp}
\phi_\perp(x) &= \int \frac{d^2 p_\perp}{16\pi^3} \; \psi_\perp (x,p_\perp^2) \nonumber \\
&=
\frac{3 N_c N_\perp M_q^2}{8\pi^2 f_v f_v^\perp} \nonumber \\
&
\times \left[ \int_{-1}^{1-2x}\mathrm{d}w \int_{\frac{2x}{w-1}+1}^{1} \mathrm{d}\beta+ \int_{1-2x}^1\mathrm{d}w \int_{\frac{2x + w-1}{w+1}}^{1} \mathrm{d}\beta \right] \nonumber \\
&
\times \frac{(1-\beta)^{\nu-1} \tilde{\rho}_v^\nu(w)}{\Omega_v^{2\nu}} \,.
\end{align}

Similarly to the pseudo-scalar meson case reported in Ref.~\cite{Albino:2022gzs}, one can conclude from Eqs.~\eqref{eq:LFWF_para} to~\eqref{eq:PDA_perp} that enticingly simple algebraic relations between the LFWFs and PDAs of a vector meson continue to exist for the transverse component:
\begin{equation}
\label{eq:AlgebraicPDA-T}
\psi_\perp(x,p_\perp^2) = \frac{16\pi^2 \nu \Omega_v^{2\nu}}{(p_\perp^2 + \Omega_v^2)^{\nu+1}} \, \phi_\perp(x) \,.
\end{equation}
However, it is no longer the case for the longitudinal one, primarily because the integrals are non-trivial and inhibit this simplification. 

The difference that exists when dealing with vector mesons instead of pseudoscalar ones comes basically from the Dirac structure of their BSA, \emph{i.e.} something unavoidable when dealing with vectors, $\gamma_\mu^T$, instead of pseudoscalars, $\gamma_5$, in a quantum field theory approach. We are interested in analytical expressions that relate the LFWF with the PDA of a particular meson; in order to do so, there is a point at which one must integrate over the transverse degrees-of-freedom, $p_\perp$. For the case of pseudoscalar mesons, one has essentially the following integral over $p_\perp$ (Eq.~(19) of Ref.~\cite{Albino:2022gzs}):
\begin{equation}
\frac{1}{8\pi^3} \int dp_\perp p_\perp  \left[ \frac{1}{(p_\perp^2+\Lambda_{1-2\alpha}^2)^{\nu+1}} \right] \,,
\end{equation}
which, for $\nu=1$, is ultraviolet finite and therefore there is nothing to remedy. In the case of vector mesons, the equivalent integral reads as:
\begin{align}
\frac{1}{8\pi^3} \int dp_\perp p_\perp \left[ \frac{1}{(p_\perp^2+\Omega_v^2)^{\nu+1}} \right] \,,
\end{align}
for the transverse part (see Eq.~\eqref{eq:LFWF_para}), which is again UV finite when $\nu=1$ and thus a similar procedure can be carried out in this case as highlighted in Eq.~\eqref{eq:AlgebraicPDA-T}. For the longitudinal part, however, we have (see Eq.~(22) of the manuscript):
\begin{align}
\frac{1}{8\pi^3} \int dp_\perp p_\perp & \Bigg[ \frac{M_q^2 + p_\perp^2 + x(1-x)m_v^2}{\left( p_\perp^2 + \Omega_v^2 \right)^{\nu+1}} \nonumber \\
&
+ \frac{1}{2\nu} \frac{1}{\left( p_\perp^2 + \Omega_v^2 \right)^{\nu}} \Bigg]\,,
\end{align}
which is logarithmically divergent in $p_\perp$ when $\nu=1$ and thus some kind of regularization procedure must be adopted. There are many ways to proceed. We have followed the one widely used within similar approaches~\cite{Raya:2015gva, Chang:2013nia}, and performed the change of the variable $\nu$ by $1+\delta$ keeping $\delta$ as a small value. This is because $\delta$ is considered an anomalous dimension~\cite{Chang:2013nia}. Therefore, it should be close to the value that mimics the ultraviolet power law behavior superimposed by logarithmic corrections which stem from the anomalous dimensions and are typical of QCD-based asymptotic behavior of the vector's BSA~\cite{Roberts:1994dr}. In fact, the value of $\delta$ in our case is $0.01$; that is to say, the new value of $\nu$ is just $1\%$ larger than the one which produces the power law part of the ultraviolet asymptotic behavior of the BSA.

We now move on to compute the SDF, required to unravel the longitudinal case.

\section{Spectral density functions}
\label{sec:SDFs}

The valence-quark (twist-two) PDAs of vector mesons, from light to heavy quark sectors, have been computed elsewhere using a rainbow-ladder truncation of QCD's DSEs~\cite{Gao:2014bca,Ding:2015rkn}. Our goal here is to derive the SDFs from such results in order to obtain the corresponding LFWFs at a later stage.

The parallel- and transverse-PDA are defined through Eqs.~\eqref{eq:PDA_para} and~\eqref{eq:PDA_perp}, respectively. Focusing on $\phi_\parallel(x)$, we perform the integral over $\beta$ and carry out the convenient shift $x\to \frac{1}{2}(1-y)$, to rewrite it as
\begin{align}
\phi_{\parallel}(y) &=  \frac{N_c N_\parallel M_q}{(\delta+1)8\pi^2 m_v f_v^2} \Bigg[ (1-y)^{\delta+1} \int_{-1}^{y}\mathrm{d}w  \frac{\tilde{\rho}_v^{\delta +1}(w)}{(1-w)^{\delta+1}} \nonumber \\
&
+ (1+y)^{\delta+1}  \int_{y}^1\mathrm{d}w \frac{\tilde{\rho}_v^{\delta+1}(w)}{(1+w)^{\delta+1}} \Bigg]  \nonumber \\
&
\times\left[\frac{M_q^2 + \frac{1}{4}(1-y^2)m_v^2}{\left[\tilde{\Omega}_v^2(y)\right]^{\delta+1}} + \frac{3}{2\delta} \frac{1}{\left[\tilde{\Omega}_v^2(y)\right]^{\delta}} \right] \,,
\end{align}
with $\tilde{\Omega}_v^2(y) = M_q^2 - \frac{1}{4}(1-y^2) m_v^2$. Proceeding in the same way for transverse-PDA, one obtains
\begin{align}
\phi_{\perp}(y) &= \frac{3N_c N_\perp}{8\pi^2 f_v f_v^{\perp} \nu} \Bigg[ (1-y)^{\delta+1} \int_{-1}^{y}\mathrm{d}w  \frac{\tilde{\rho}_v^{\delta +1} (w)}{(1-w)^{\delta+1}} \nonumber \\
&
+ (1+y)^{\delta +1}  \int_{y}^1\mathrm{d}w \frac{\tilde{\rho}_v^{\delta +1} (w)}{(1+w)^{\delta+1}} \Bigg]  \nonumber \\
&
\times \left[\frac{M_q^2}{\left[\tilde{\Omega}_v^2(y)\right]^{\nu}} \right] \,.
\end{align}
The PDAs can be rewritten as
\begin{equation}
\phi_{\parallel(\perp)}(y) = \varphi_{\parallel(\perp)}^{-}(y) + \varphi_{\parallel(\perp)}^{+}(y) \,,
\label{defPDAseparation}
\end{equation}
where we have used the definition
\begin{align}
\varphi_{\parallel(\perp)}^{\pm}(y) &=\mp\, \tau_{\parallel(\perp)}^{\pm}(y) \int_{\pm 1}^y  \frac{\tilde{\rho}_v^{\delta +1} (w)}{(1\pm w)^{\delta+1}} \label{defvarphi} \,,
\end{align}
with
\begin{align}
\tau_{\parallel}^{\pm}(y) &= \frac{N_c N_\parallel M_q (1\pm y)^{\delta+1}}{(\delta+1)8\pi^2 m_v f_v^2} \Bigg\{ \frac{M_q^2 + \frac{1}{4}(1-y^2)m_v^2}{\left[\tilde{\Omega}_v^2(y)\right]^{\delta+1}} \nonumber \\
&
+ \frac{3}{2\delta} \frac{1}{\left[\tilde{\Omega}_v^2(y)\right]^{\delta}} \Bigg\}, \\
\tau_{\perp}^{\pm}(y) &= \frac{3 N_c N_\perp M_q^2 (1\pm y)^{\delta+1}}{8\pi^2 f_v f_v^{\perp}\nu} \frac{1}{\left[\tilde{\Omega}_v^2(y)\right]^{\nu}} \,.
\end{align}

It is possible to compute an expression for the SDF in terms of the derivatives of the corresponding PDAs. To do this, we calculate (for simplicity of notation we omit the subscripts $\parallel$ and $\perp$ from now on):
\begin{align}
\frac{\mathrm{d}\phi(y)}{\mathrm{d}y} =&\; \frac{\mathrm{d}\varphi^- (y)}{\mathrm{d}y} + \frac{\mathrm{d}\varphi^+(y)}{\mathrm{d}y} \label{derPDA} \,, \\
\frac{\mathrm{d}^2\phi(y)}{\mathrm{d}y^2} =&\; \frac{\mathrm{d}^2\varphi^- (y)}{\mathrm{d}y^2} + \frac{\mathrm{d}^2\varphi^+(y)}{\mathrm{d}y^2} \,. \label{der2PDA}
\end{align}
Differentiating the expression in Eq.~\eqref{defvarphi} with respect to the variable $y$, we obtain
\begin{align}\label{dervarphi}
\frac{\mathrm{d}\varphi^{\pm}(y)}{\mathrm{d}y} &= \mp\, \tau^{\pm}(y) \frac{\mathrm{d}}{\mathrm{d}y}\int_{\pm1}^y  \frac{\tilde{\rho}_v^{\delta +1} (w)}{(1\pm w)^{\delta+1}} \nonumber \\
&
\mp\, \frac{\mathrm{d}\tau^{\pm}(y)}{\mathrm{d}y}\int_{\pm1}^y  \frac{\tilde{\rho}_v^{\delta +1} (w)}{(1\pm w)^{\delta+1}} \nonumber \\   
&
= \mp\, \tau^{\pm}(y) \frac{\tilde{\rho}_v^{\delta +1} (y)}{(1\pm y)^{\delta+1}} \nonumber \\
&
\mp\, \frac{\mathrm{d}\tau^{\pm}(y)}{\mathrm{d}y}\int_{\pm1}^y  \frac{\tilde{\rho}_v^{\delta +1} (w)}{(1\pm w)^{\delta+1}} \,,   
\end{align}
where the fundamental theorem of calculus has been applied in the last line. Now using Eq.~\eqref{defvarphi}, the expression above can be rearranged as
\begin{align}\label{dervarphi2}
&
\frac{\mathrm{d}\varphi^{\pm}(y)}{\mathrm{d}y} = \mp \, \tau^{\pm}(y) \frac{\tilde{\rho}_v^{\delta +1}(y)}{(1\pm y)^{\delta+1}} + \frac{\varphi^{\pm}(y)}{\tau^{\pm}(y)}\frac{\mathrm{d}\tau^{\pm}(y)}{\mathrm{d}y} \nonumber \\
&
= \mp \, \left[\frac{\Lambda_y^2}{1\pm y}\right]^{\delta+1} \rho_v(y) \tau^{\pm}(y) + \frac{\varphi^{\pm}(y)}{\tau^{\pm}(y)} \frac{\mathrm{d}\tau^{\pm}(y)}{\mathrm{d}y} \,,
\end{align}
where, again, $\Lambda_y^2=\Lambda^2(y)$ and $\tilde{\rho}_v^{\delta+1}(y) = \Lambda_y^{2(\delta+1)} \rho_v(y)$. 
We notice that the first term of Eq.~\eqref{dervarphi} vanishes in the sum of Eq.~\eqref{derPDA}, as well as in Eq.~\eqref{der2PDA}; so it can safely be neglected. Repeating the step and differentiating Eq.~\eqref{dervarphi} with respect to $y$,
\begin{align}\label{der2varphi}
&
\frac{\mathrm{d}^2\varphi^{\pm}(y)}{\mathrm{d}y^2} = \mp \, \frac{\mathrm{d}\tau^{\pm}(y)}{\mathrm{d}y} \frac{\mathrm{d}}{\mathrm{d}y} \int_{\pm1}^y  \frac{\tilde{\rho}_v^{\delta +1}(w)}{(1\pm w)^{\delta+1}} \nonumber \\
&
\mp \, \frac{\mathrm{d}^2\tau^{\pm}(y)}{\mathrm{d}y^2} \int_{\pm1}^y  \frac{\tilde{\rho}_v^{\delta +1} (w)}{(1\pm w)^{\delta+1}} \nonumber \\   
&
= \mp \, \frac{\mathrm{d}\tau^{\pm}(y)}{\mathrm{d}y} \frac{\tilde{\rho}_v^{\delta +1} (y)}{(1\pm y)^{\delta+1}} +  \frac{\varphi^{\pm}(y)}{\tau^{\pm}(y)} \frac{\mathrm{d}^2\tau^{\pm}(y)}{\mathrm{d}y^2} \nonumber \\
&
= \mp \, \frac{\mathrm{d}\tau^{\pm}(y)}{\mathrm{d}y} \left[\frac{\Lambda_y^2}{1\pm y}\right]^{\delta+1} \rho_v(y) +   \frac{\varphi^{\pm}(y)}{\tau^{\pm}(y)} \frac{\mathrm{d}^2\tau^{\pm}(y)}{\mathrm{d}y^2} \,.
\end{align}
From Eq.~\eqref{defPDAseparation} we have that
\begin{equation}\label{varphiminus}
\varphi^-(y) = \phi(y) - \varphi^+(y) \,;
\end{equation}
inserting Eqs.~\eqref{dervarphi2} and~\eqref{varphiminus} into~\eqref{derPDA}, one readily arrives at the result:
\begin{align}
&
\frac{\mathrm{d}\phi(y)}{\mathrm{d}y} = \varphi^{+}(y) \left[ \frac{1}{\tau^{+}(y)}\frac{\mathrm{d}\tau^{+} (y)}{\mathrm{d}y} - \frac{1}{\tau^{-}(y)}\frac{\mathrm{d}\tau^{-} (y)}{\mathrm{d}y}   \right] \nonumber \\
&
+ \rho_v (y) \left[ \left(\frac{\Lambda^2_y}{1-y}\right)^{\delta + 1} \tau^{-}(y) - \left(\frac{\Lambda^2_y}{1+y}\right)^{\delta + 1} \tau^{+}(y)\right] \nonumber \\
&
+ \frac{\mathrm{d}\tau^{-}(y)}{\mathrm{d}y} \frac{\phi(y)}{\tau^{-}(y)} \,.
\end{align}
One can solve this for $\varphi^{+}(y)$ to obtain:
\begin{align}
\varphi^{+}(y) &= \left[ \frac{1}{\tau^{+}(y)}\frac{\mathrm{d}\tau^{+} (y)}{\mathrm{d}y} - \frac{1}{\tau^{-}(y)}\frac{\mathrm{d}\tau^{-} (y)}{\mathrm{d}y}   \right]^{-1} \nonumber \\
&
\times \Bigg\{  \frac{\mathrm{d}\phi(y)}{\mathrm{d}y} - \frac{\mathrm{d}\tau^{-}(y)}{\mathrm{d}y} \frac{\phi(y)}{\tau^{-}(y)} \nonumber \\
& 
- \Bigg[ \left(\frac{\Lambda^2_y}{1-y}\right)^{\delta + 1} \tau^{-}(y) \nonumber \\
&
\hspace*{0.80cm} - \left(\frac{\Lambda^2_y}{1+y}\right)^{\delta + 1} \tau^{+}(y) \Bigg] \rho_v (y) \Bigg\} \,.
\end{align}
On substituting $\tau^{\pm}$, the last terms vanish, leaving us with
\begin{align}\label{varphiplus}
\varphi^{+}(y) &= \left[ \frac{1}{\tau^{+}(y)}\frac{\mathrm{d}\tau^{+} (y)}{\mathrm{d}y} - \frac{1}{\tau^{-}(y)}\frac{\mathrm{d}\tau^{-} (y)}{\mathrm{d}y} \right]^{-1} \nonumber \\
&
\times \left[  \frac{\mathrm{d}\phi(y)}{\mathrm{d}y} - \frac{\mathrm{d}\tau^{-}(y)}{\mathrm{d}y} \frac{\phi(y)}{\tau^{-}(y)} \right] \,.
\end{align}
Now, inserting Eqs.~\eqref{der2varphi} and~\eqref{varphiminus} into~\eqref{der2PDA}, we get 
\begin{align}
&
\frac{\mathrm{d}^2\phi(y)}{\mathrm{d}y^2} = \varphi^{+}(y) \left[ \frac{1}{\tau^{+}(y)}\frac{\mathrm{d}^2\tau^{+} (y)}{\mathrm{d}y^2} - \frac{1}{\tau^{-}(y)}\frac{\mathrm{d}^2\tau^{-} (y)}{\mathrm{d}y^2}   \right] \nonumber \\
&
+ \rho_v (y) \left[ \left(\frac{\Lambda^2_y}{1-y}\right)^{\delta + 1} \frac{\mathrm{d}\tau^{-}(y)}{\mathrm{d}y} - \left(\frac{\Lambda^2_y}{1+y}\right)^{\delta + 1} \frac{\mathrm{d}\tau^{+}(y)}{\mathrm{d}y}\right] \nonumber \\
&
+ \frac{\mathrm{d}^2\tau^{-}(y)}{\mathrm{d}y^2} \frac{\phi(y)}{\tau^{-}(y)} \,.
\end{align}
Substituting Eq.~\eqref{varphiplus} in the expression above and solving for $\rho_v(y)$, one obtains an expression for the SDF in terms of the (leading-twist) PDA:
\begin{align}
\rho_v (y) &= \frac{1}{A(y)} \Bigg[ \frac{\mathrm{d}^2\phi(y)}{\mathrm{d}y^2} - \frac{\mathrm{d}^2\tau^{-}(y)}{\mathrm{d}y^2} \frac{\phi(y)}{\tau^{-}(y)} \nonumber \\
&
- \frac{B(y)}{C(y)} \left( \frac{\mathrm{d}\phi(y)}{\mathrm{d}y} - \frac{\mathrm{d}\tau^{-}(y)}{\mathrm{d}y} \frac{\phi(y)}{\tau^{-}(y)} \right) \Bigg] \,,
\end{align}
with
\begin{subequations}
\begin{align}
A(y) &=  \left(\frac{\Lambda^2_y}{1-y}\right)^{\delta + 1} \frac{\mathrm{d}\tau^{-}(y)}{\mathrm{d}y} \nonumber \\
&
- \left(\frac{\Lambda^2_y}{1+y}\right)^{\delta + 1} \frac{\mathrm{d}\tau^{+}(y)}{\mathrm{d}y} \,, \\
B(y) &=  \frac{1}{\tau^{+}(y)}\frac{\mathrm{d}^2\tau^{+} (y)}{\mathrm{d}y^2} - \frac{1}{\tau^{-}(y)}\frac{\mathrm{d}^2\tau^{-} (y)}{\mathrm{d}y^2} \,, \\
C(y) &= \frac{1}{\tau^{+}(y)}\frac{\mathrm{d}\tau^{+} (y)}{\mathrm{d}y} - \frac{1}{\tau^{-}(y)}\frac{\mathrm{d}\tau^{-} (y)}{\mathrm{d}y} \,.
\end{align}
\end{subequations}
This is the final expression for the SDF. We now have all the necessary tools to compute the LFWF in terms of the PDAs within this algebraic model.


\section{Results}
\label{sec:results}

The main goal of this manuscript is to provide a simple relation between the PDAs and their corresponding LFWFs of hidden-flavor vector mesons, from light to heavy quark sectors. Our starting point was the recent development of a similar theoretical framework for the lowest-lying hidden-flavor pseudo-scalar mesons in Ref.~\cite{Albino:2022gzs}. Such a connection has been derived, developed and presented above, concluding that the transverse LFWFs can be obtained directly from the corresponding PDAs through the algebraic relation presented in Eq.~\eqref{eq:AlgebraicPDA-T}, whereas the calculation of longitudinal LFWFs requires an intermediate step where the SDF must be derived from the particular PDA. We are now in the position to provide numerical results for the LFWFs from the most up-to-date knowledge of the corresponding PDAs.

A rainbow-ladder (RL) truncation of QCD's DSEs, defined by an interaction compatible with modern studies of the gauge sector~\cite{Aguilar:2012rz, Ayala:2012pb, Aguilar:2013hoa, Binosi:2014aea, Aguilar:2019uob, Zafeiropoulos:2019flq}, was used in Ref.~\cite{Gao:2014bca} to compute $\rho$- and $\phi$-meson valence-quark (twist-two) PDAs. These results are faithfully described by the following simple parametrizations (renormalization scale $\xi_2=2\,\text{GeV}$):
\begin{subequations}
\begin{align}
\phi_{\parallel}^{\rho}(\xi_2;x) &= 3.26 \, x^{0.66} \, (1-x)^{0.66} \,, \\
\phi_{\perp}^{\rho}(\xi_2;x) &= 2.73 \, x^{0.49} \, (1-x)^{0.49} \,, \\
\phi_{\parallel}^{\phi}(\xi_2;x) &= 3.14 \, x^{0.64} \, (1-x)^{0.64} \,, \\
\phi_{\perp}^{\phi}(\xi_2;x) &= 2.64 \, x^{0.48} \, (1-x)^{0.48} \,,
\end{align}
\end{subequations}
or equivalently, through following expressions exhibiting explicitly soft end-point behavior
\begin{subequations}
\begin{align}
\phi_{\parallel}^{\rho}&(\xi_2;x) = 15.43x(1-x) \nonumber \\
&
\times \Bigg[ 1 + 1.84x(1-x) - 2.22 \sqrt{x(1-x)} \Bigg] \,, \\
\phi_{\perp}^{\rho}&(\xi_2;x) = 22.26x(1-x) \nonumber \\
&
\times \Bigg[ 1 + 2.44x(1-x) - 2.76\sqrt{x(1-x))} \Bigg] \,, \\
\phi_{\parallel}^{\phi}&(\xi_2;x) = 16.15x(1-x) \nonumber \\
&
\times \Bigg[ 1 + 1.92x(1-x) - 2.29\sqrt{x(1-x)} \Bigg] \,, \\
\phi_{\perp}^{\phi}&(\xi_2;x) = 23.77x(1-x) \nonumber \\
&
\times \Bigg[ 1 + 2.61x(1-x) - 2.88\sqrt{x(1-x)} \Bigg] \,.
\end{align}
\end{subequations}

The RL truncation has also been explored in connection with heavy-light and heavy-heavy mesons~\cite{Bender:2002as, Bhagwat:2004hn, Nguyen:2010yh, Blank:2011ha, Rojas:2014aka, Hilger:2014nma}. The main conclusion of those studies was that corrections which go beyond the RL-truncation of the dressed--quark-gluon vertex and hence the Bethe-Salpeter kernel are critical in heavy-light systems; and an interaction strength for the RL kernel fitted to pion properties alone is not optimal in the treatment of these systems.

Following the breakthrough progress in Ref.~\cite{Chang:2009zb}, it has become possible to employ more sophisticated kernels for the quark propagator DSE and BSEs, which overcome the weaknesses of RL truncation in all channels studied thus far and, in particular, within the heavy quark sectors~\cite{Yin:2019bxe, Yin:2021uom, Qin:2019hgk}. This new symmetry preserving technique has the key feature that it expresses more DCSB in the integral equations connected with bound-states, and less in the DSE that describes the quark propagator, arriving at a more balanced description than the RL-truncation when solving equations involving different mass scales simultaneously. Once the kernels have been specified, one can employ standard algorithms, and approximations, to obtain numerical solutions for the PDAs of the lowest-lying vector quarkonia~\cite{Ding:2015rkn}, whose parametrizations can be given by
\begin{subequations}
\begin{align}
\phi_{\parallel(\perp)}^{J/\psi(\Upsilon(1S))}(\xi_2;\alpha=2x-1) &= \frac{3}{2} N(a) (1-\alpha^2) e^{-\alpha^2 a^2} \,,
\end{align}
\end{subequations}
where
\begin{equation}
N(a) = \frac{8a^3}{3\sqrt{\pi} (2a^2-1) \erf(a) + 6 a e^{-a^2}} \,,
\end{equation}
with
\begin{subequations}
\begin{align}
a_{\parallel}^{J/\psi} &= 3.4 \,, \hspace*{0.95cm} a_{\perp}^{J/\psi} = 3.0 \,, \\
a_{\parallel}^{\Upsilon(1S)} &= 6.0 \,, \quad\quad a_{\perp}^{\Upsilon(1S)} = 5.3 \,.
\end{align}
\end{subequations}
and $\text{erf}(z)=2/\sqrt{\pi}\int_0^z e^{-t^2} dt$ is the Gauss error function.

\begin{figure}[!t]
\centerline{\includegraphics[width=0.5\textwidth]{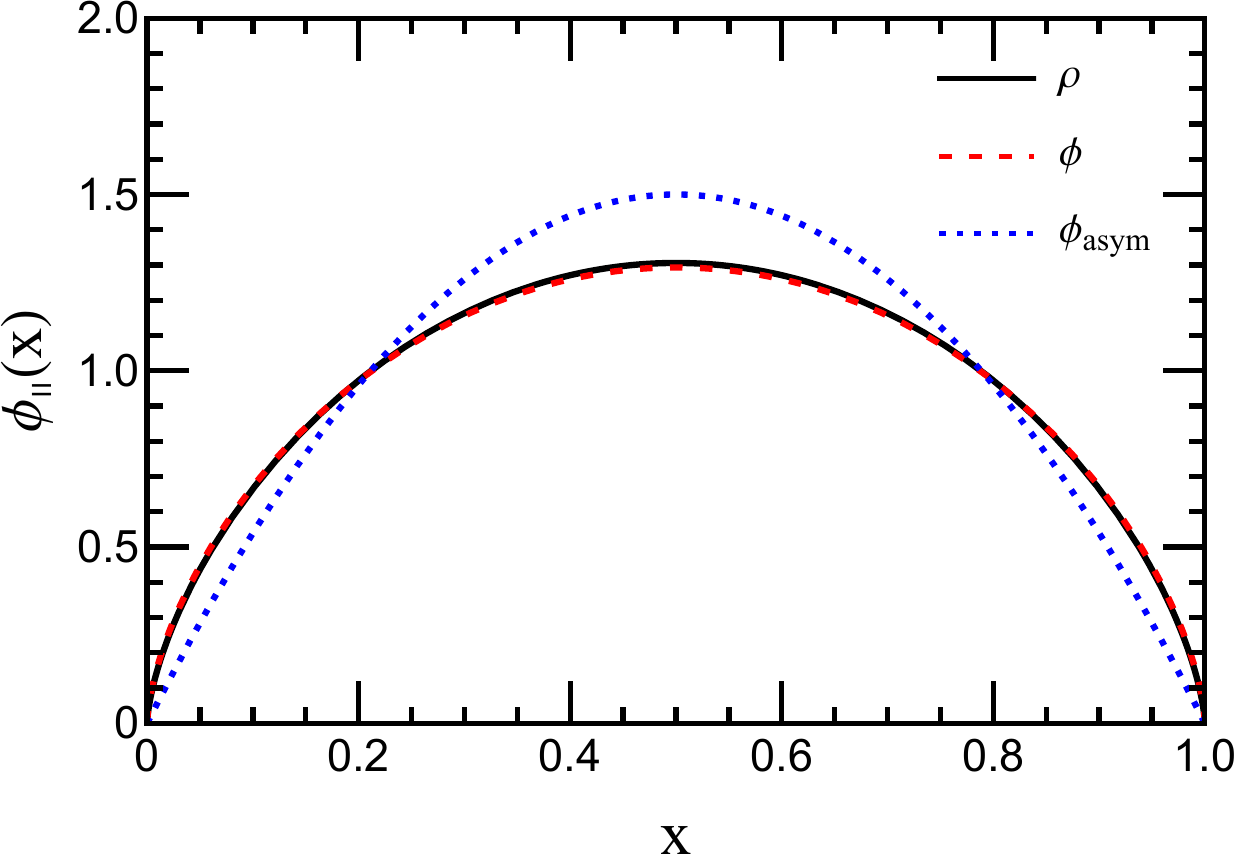}}
\vspace*{0.50cm}
\centerline{\includegraphics[width=0.5\textwidth]{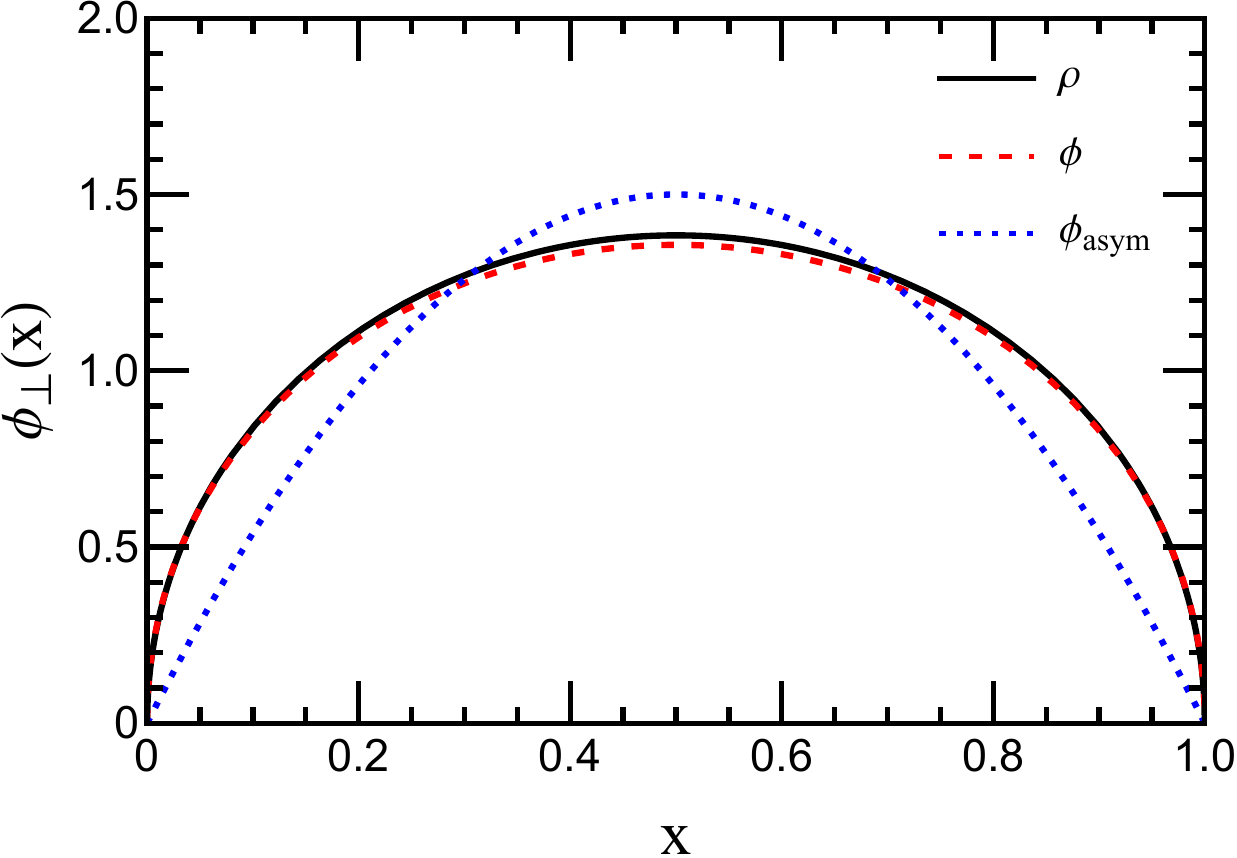}}
\caption{\label{fig:PDA-light} Valence-quark (twist-two) PDAs for the $\rho$- and the $\phi$-meson at an energy scale $\xi_2=2\,\text{GeV}$. \emph{Upper panel:} longitudinal PDAs. \emph{Lower panel:} transverse PDAs. The legend of the curves is the following: \emph{black solid}, $\phi_{\parallel(\perp)}^\rho(x)$; \emph{red dashed}, $\phi_{\parallel(\perp)}^\phi(x)$; and \emph{blue dotted}, $\varphi^{\rm asy}(x)=6x(1-x)$, \emph{i.e.} the PDA associated with QCD's conformal limit~\cite{Lepage:1979zb, Efremov:1979qk, Lepage:1980fj}.}
\end{figure}

\begin{figure}[!t]
\centerline{\includegraphics[width=0.49\textwidth]{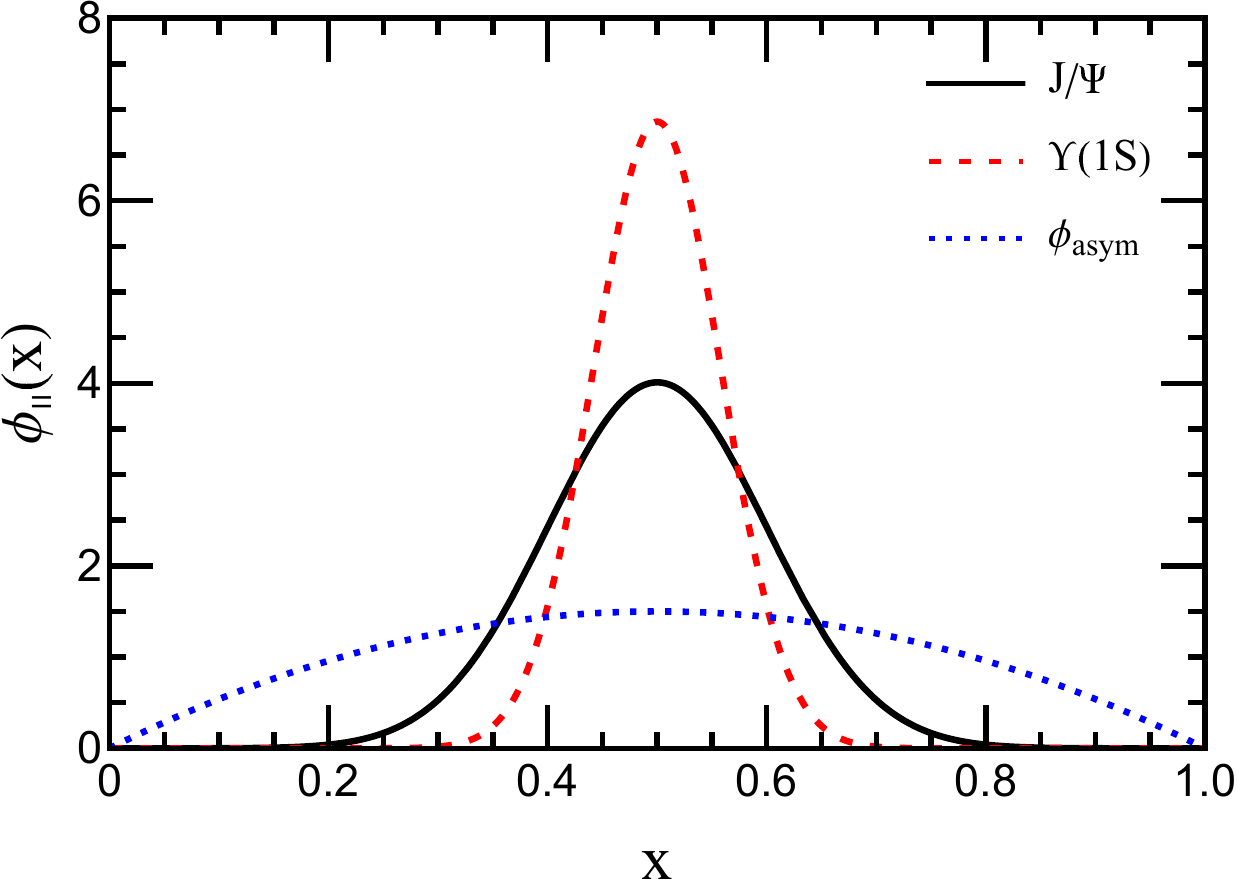}}
\vspace*{0.50cm}
\centerline{\includegraphics[width=0.49\textwidth]{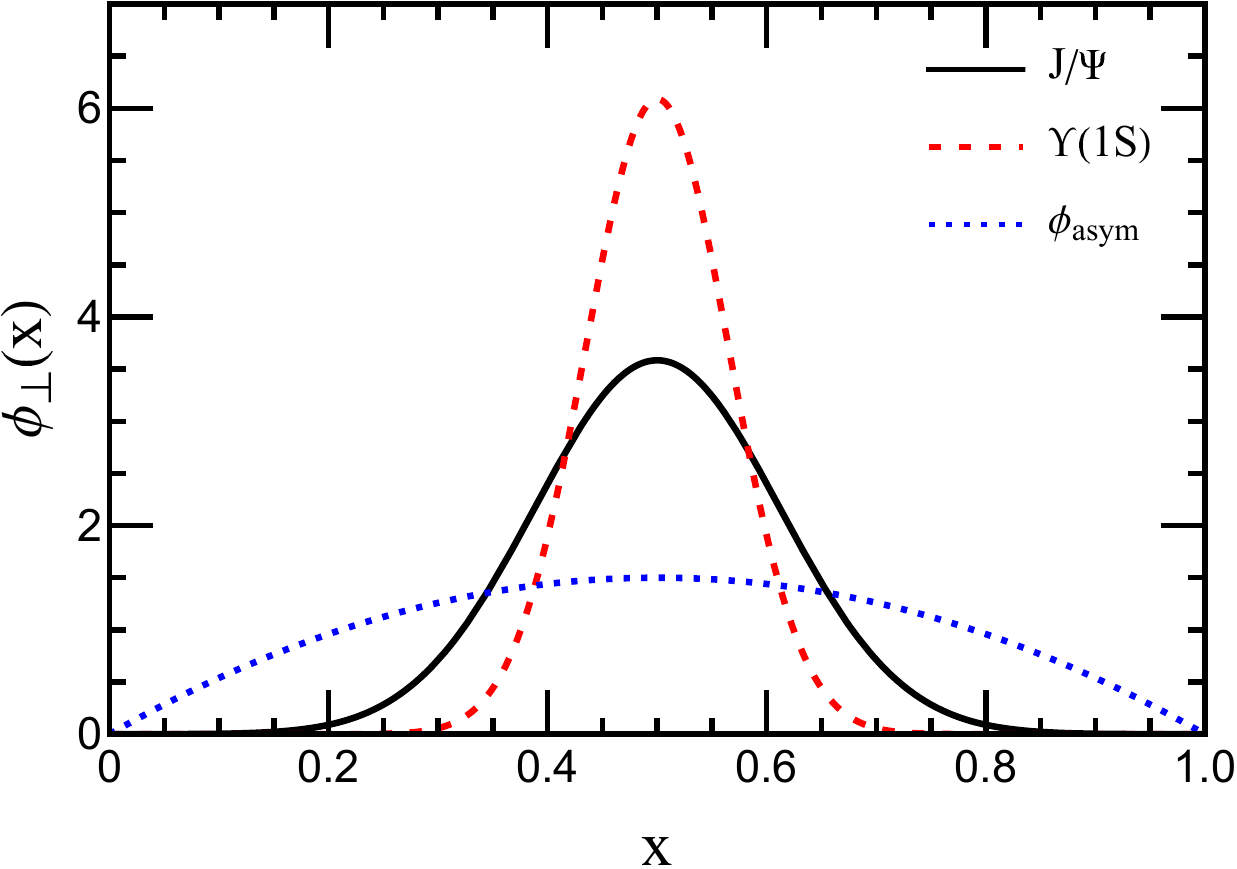}}
\caption{\label{fig:PDA-heavy} Valence-quark (twist-two) PDAs for the $J/\psi$- and $\Upsilon(1S)$-meson at an energy scale $\xi_2=2\,\text{GeV}$. \emph{Upper panel:} longitudinal PDAs. \emph{Lower panel:} transverse PDAs. The legend of the curves is the following: \emph{black solid}, $\phi_{\parallel(\perp)}^{J/\psi}(x)$; \emph{red dashed}, $\phi_{\parallel(\perp)}^{\Upsilon(1S)}(x)$; and \emph{blue dotted}, $\varphi^{\rm asy}(x)=6x(1-x)$, \emph{i.e.} the PDA associated with QCD's conformal limit~\cite{Lepage:1979zb, Efremov:1979qk, Lepage:1980fj}.}
\end{figure}

Figure~\ref{fig:PDA-light} shows the point-wise behavior of the $\rho$- and $\phi$-meson PDAs. Accumulated evidence suggests that the \linebreak PDAs associated with light-quark vector meson charge-conjugation eigenstates are concave functions whose widths are ordered as follows
\begin{equation}
\label{sizemeson}
\phi^{\text{asy}} <_{N} \phi_{\parallel}^{\rho} <_{N} \phi_{\parallel}^{\phi} <_{N} \phi_\perp^{\rho} <_{N} \phi_\perp^{\phi} \,,
\end{equation}
where ``$<_{N}$'' means ``narrower than''. The inequality~\eqref{sizemeson} actually signals that each vector meson PDA is significantly broader than the asymptotic distribution, a feature strongly related with DCSB. Within the context of QCD, a useful measure of this dilation is the energy scale to which one must evolve a given PDA in order that $\phi^{\rm asy}$ may be considered a reliable approximation. As remarked in Ref.~\cite{Gao:2014bca}, it is not uniformly achieved for the vector mesons unless $\zeta \gtrsim \zeta_{\rm LHC}$, which is the energy accessible at the LHC.

Figure~\ref{fig:PDA-heavy} shows the $x$-dependence of the $J/\psi$- and $\Upsilon(1S)$-meson PDAs. It is evident that the PDAs associated with heavy-quark meson charge-conjugation eigenstates are piecewise convex-concave-convex, showing the following ordering
\begin{equation}
\phi_{\Upsilon}^\| <_N \phi_{\Upsilon}^\bot <_N \phi_{J/\Psi}^\| <_N \phi_{J/\Psi}^\bot <_N \phi^{\text{asy}} \,.
\end{equation}
This is to say, the PDAs of the lowest-lying heavy vector quarkonia are much narrower than $\phi^{\rm asy}$ on a large domain of the energy scale, $\xi$, and they naturally evolve to the fixed point $\phi^{\text{asy}}$ as $\xi$ is increased. Nevertheless, for realistic values of either charm or bottom quark mass, the PDAs deviate noticeably from $\phi(x) = \delta(x-1/2)$, which is the limiting form in the case of infinitely heavy (static) quarks.

At this point, it is worth emphasizing that all existing continuum quantum field theory studies of meson valence-quark distribution functions (DFs) begin with expressions of the same kind as Eqs.~\eqref{eq:PDAs}. Therefore, in a computational framework that preserves the multiplicative renormalizability of QCD, one has
\begin{equation}
S_q(\xi_1;p) \Gamma_\mu^v(\xi_1;p,P) = S_q(\xi_2;p) \Gamma_\mu^v(\xi_2;p,P) \,,
\end{equation}
implying that $\phi(\xi_1;x)=\phi(\xi_2;x)$, \emph{i.e.} the result is actually independent of the renormalization point as long as it is expressed in a quasi-particle basis~\cite{Cui:2020tdf}. This characteristic is good so far as baryon number conservation is concerned because it guarantees
\begin{equation}
1 = \int_0^1 dx\, \phi(\xi_1;x) = \int_0^1 dx\, \phi(\xi_2;x) \,.
\end{equation}
However, it also imposes the same identity for all $n\geq 1$ moments, \emph{i.e.} it precludes DF evolution. Consequently, the PDAs expressed as Eqs.~\eqref{eq:PDAs} and depicted in Figs.~\ref{fig:PDA-light} and~\ref{fig:PDA-heavy} must be interpreted as valid for $\xi = \xi_H \approx 0.33\,\text{GeV}$, intuitively identified with some hadronic scale for which the valence degrees of freedom fully express the properties of the hadron under study.

The quark and antiquark PDAs are connected via momentum conservation,
\begin{eqnarray}
\label{eq:PDAquarkantiquark}
\phi_{v}^q(\xi_H;x)=\phi_{v}^{\bar q}(\xi_H;1-x) \;,
\end{eqnarray}
a constricted and firm connection that prevails even after evolution~\cite{Lepage:1979zb, Efremov:1979qk, Lepage:1980fj}. Finally, note also that, for mesons with equal quark and antiquark flavors, $\phi_{\parallel,\perp}(x)$ are even under the exchange $x \leftrightarrow \bar x=(1-x)$; and they vanish at the endpoints unless the underlying interaction is momentum-independent.

It is also worth noting herein that the presented formalism is equally valid for any momentum-scale that characterizes the exclusive process in which the meson is involved. That is to say, given the valence-quark (twist-two) PDA of a vector-meson charge-conjugation eigenstate at a given momentum scale, $\xi$, the formalism provides us the corresponding leading-twist LFWF of the vector meson at the same energy scale.

Let us now proceed with the computation of the longitudinal and transverse LFWFs of the $\rho(770)$, $\phi(1020)$, $J/\psi$ and $\Upsilon(1S)$ mesons through their connection with the corresponding PDAs. The useful expressions can be found in Eqs.~\eqref{eq:LFWF_para} and~\eqref{eq:AlgebraicPDA-T}, \emph{i.e.}, while the transverse LFWF can be computed algebraically from its analogous PDA, the longitudinal LFWF must be calculated using the SDF derived from its PDA partner. Besides, the parameters needed to compute the LFWFs are shown in Table~\ref{tab:Parameters}. Note that these are not free-parameters of this analysis, these are computed in connection with the PDAs collected herein and were originally reported in Refs.~\cite{Gao:2014bca, Ding:2015rkn}. As shown in Table~\ref{tab:Parameters}, the vector meson static properties are in fairly good agreement with the experimental data collected in the Review of Particle Physics by the Particle Data Group~\cite{ParticleDataGroup:2022pth}.

\begin{table}[!t]
\caption{\label{tab:Parameters} Vector meson static properties computed using RL truncation for light mesons and DB-improved kernels for heavy mesons (see text for details).
All quantities in GeV and $f_v^\perp$ values are quoted at a renormalisation scale $\xi = \xi_2 = 2\,\text{GeV}$.
The current-quark masses are $m_{u/d}(\xi=\xi_2=2\,\text{GeV})=4.5\,\text{MeV}$, $m_{s}(\xi_2)=99\,\text{MeV}$, $m_c(\xi_2)=1.22\,\text{GeV}$, $m_b(\xi_2)=4.17\,\text{GeV}$ which correspond to the following dressed quark masses $M_{u/d}(\xi=\xi_0=0\,\text{GeV})=0.56\,\text{GeV}$, $M_s(\xi_0)=0.80\,\text{GeV}$, $M_c(\xi_0)=1.42\,\text{GeV}$, $M_b(\xi_0)=4.49\,\text{GeV}$.
The experimental data (Exp.) are taken from Ref.~\cite{ParticleDataGroup:2022pth}.}
\begin{ruledtabular}
\begin{tabular}{ccccc}
Meson & Approach & $m_v$ & $f_v$ & $f_v^\perp$ \\
\hline
$\rho(770)$    & The. & $0.740$ & $0.150$ & $0.110$ \\
               & Exp. & $0.775$ & $0.156$ & - \\[2ex]
$\phi(1020)$   & The. & $1.08$ &  $0.190$ &  $0.150$ \\
               & Exp. & $1.02$ & $0.161$ & - \\[2ex]
$J/\psi$       & The. & $3.07$ & $0.255$ & $0.213$ \\
               & Exp. & $3.10$ & $0.294$ & - \\[2ex]
$\Upsilon(1S)$ & The. & $9.46$ & $0.471$ & $0.421$ \\
               & Exp. & $9.46$ & $0.506$ & - \\
\end{tabular}
\end{ruledtabular}
\end{table}

The absolute values as well as the ratios of the vector meson decay constants reveal interesting features of these mesons related with either the dynamical or explicit breaking of chiral symmetry within the light or heavy quark sectors, respectively. This is because the vector decay constants, $f_v$, are associated with currents which are invariant under chiral transformations, whereas the currents that define the tensor decay constants, $f_v^\perp$, are not chirally invariant and hence the values of $f_v^\perp$ are a clear expression of the strength of DCSB within the vector mesons. At $\xi=2\,\text{GeV}$, we have
\begin{subequations}
\begin{align}
f_\rho^\perp / f_\rho &= 0.73 \,, \hspace*{1.55cm} f_\phi^\perp / f_\phi = 0.79 \,, \\
f_{J/\psi}^\perp / f_{J/\psi} &= 0.84 \,, \hspace*{0.50cm} f_{\Upsilon(1S)}^\perp / f_{\Upsilon(1S)} = 0.89 \,,
\end{align}
\end{subequations}
and thus, at an hadronic scale, chiral symmetry breaking is strong and dynamical within the $\rho$- and $\phi$-mesons; while it is weaker and explicit within the $J/\psi$ and $\Upsilon(1S)$ mesons.

\begin{figure}[!t]
\centerline{\includegraphics[width=0.45\textwidth]{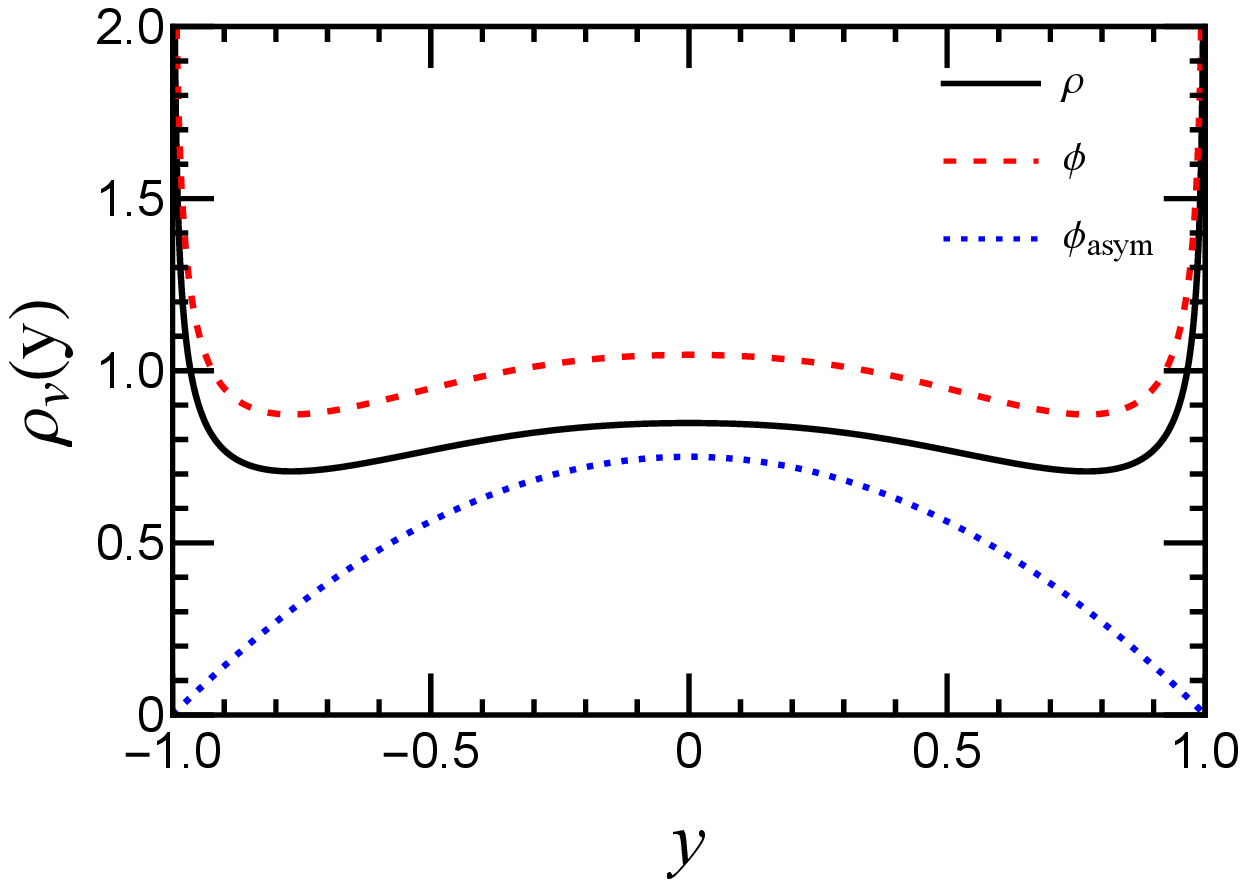}}
\vspace*{0.50cm}
\centerline{\includegraphics[width=0.45\textwidth]{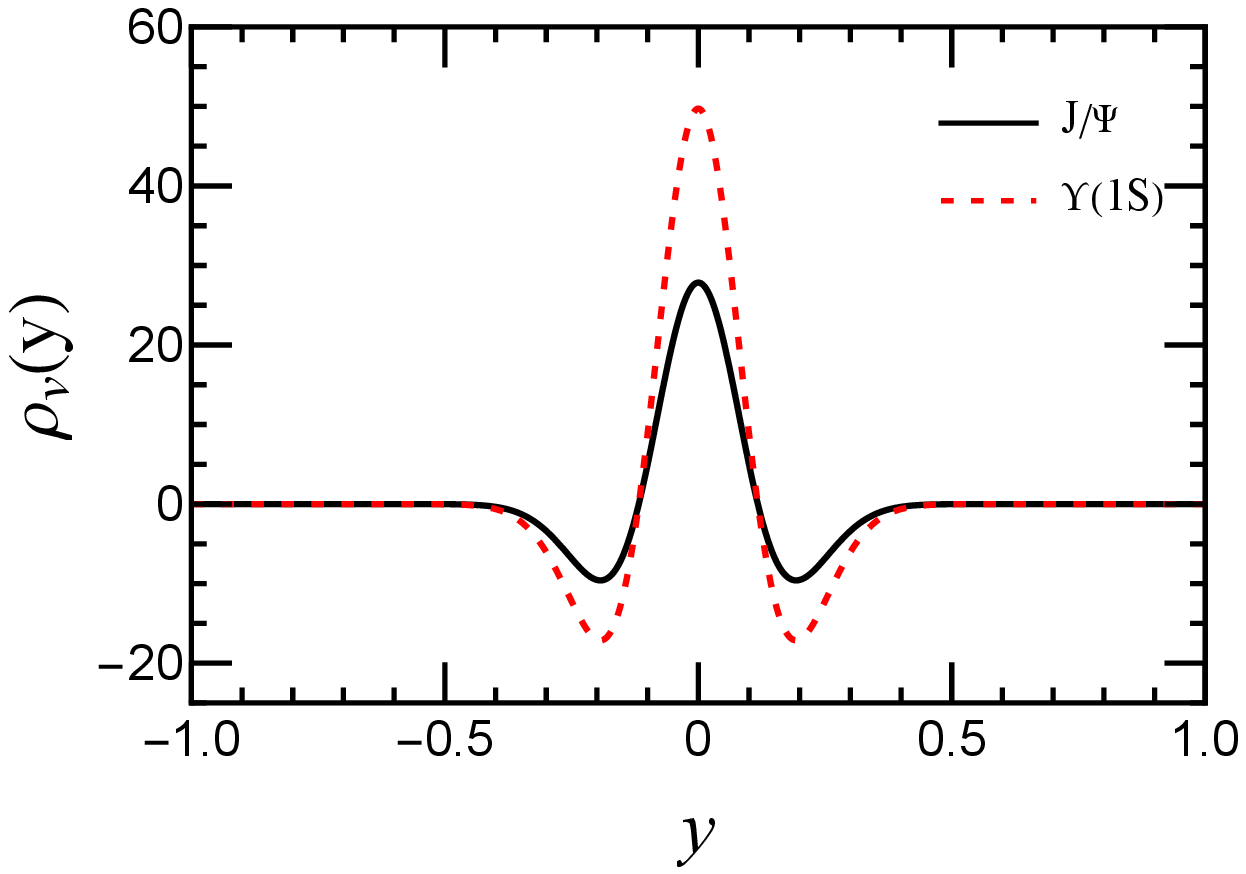}}
\caption{\label{fig:SpectralDensities} Spectral density functions for the hidden light-flavor (upper panel) and heavy-flavor (lower panel) vector mesons. For the upper panel, the blue dotted line is the spectral density function corresponding to the PDA associated with QCD's conformal limit~\cite{Lepage:1979zb, Efremov:1979qk, Lepage:1980fj}.}
\end{figure}

We now turn our attention to the SDFs of the $\rho(770)$, $\phi(1020)$, $J/\psi$ and $\Upsilon(1S)$ mesons, derived from their longitudinal PDAs. The upper panel of Fig.~\ref{fig:SpectralDensities} shows the SDFs corresponding to the hidden light-flavor vector mesons. One can see the following salient features: the $\rho$ and $\phi$ mesons present very similar SDFs; both are positive definite in the whole range, with a point-wise behavior characterized by alternating convex and concave forms which resemble the trend obtained already in Ref.~\cite{Albino:2022gzs} for the lowest-lying pseudo-scalar mesons. However, their end points are non-zero, in contrast with the pseudo-scalar case. The lower panel of Fig.~\ref{fig:SpectralDensities} shows the SDFs of hidden heavy-flavor vector mesons. One can see in this case that the SDFs are narrower than those of light vector mesons and, furthermore, they are not positive definite on the entire interval. This last feature is not a discrepancy of our theoretical approach since the only condition that must be fulfilled is that the area under the curve must be positive definite, which is actually the case.

\begin{figure*}[t!]
\centering
\includegraphics[width=0.45\textwidth]{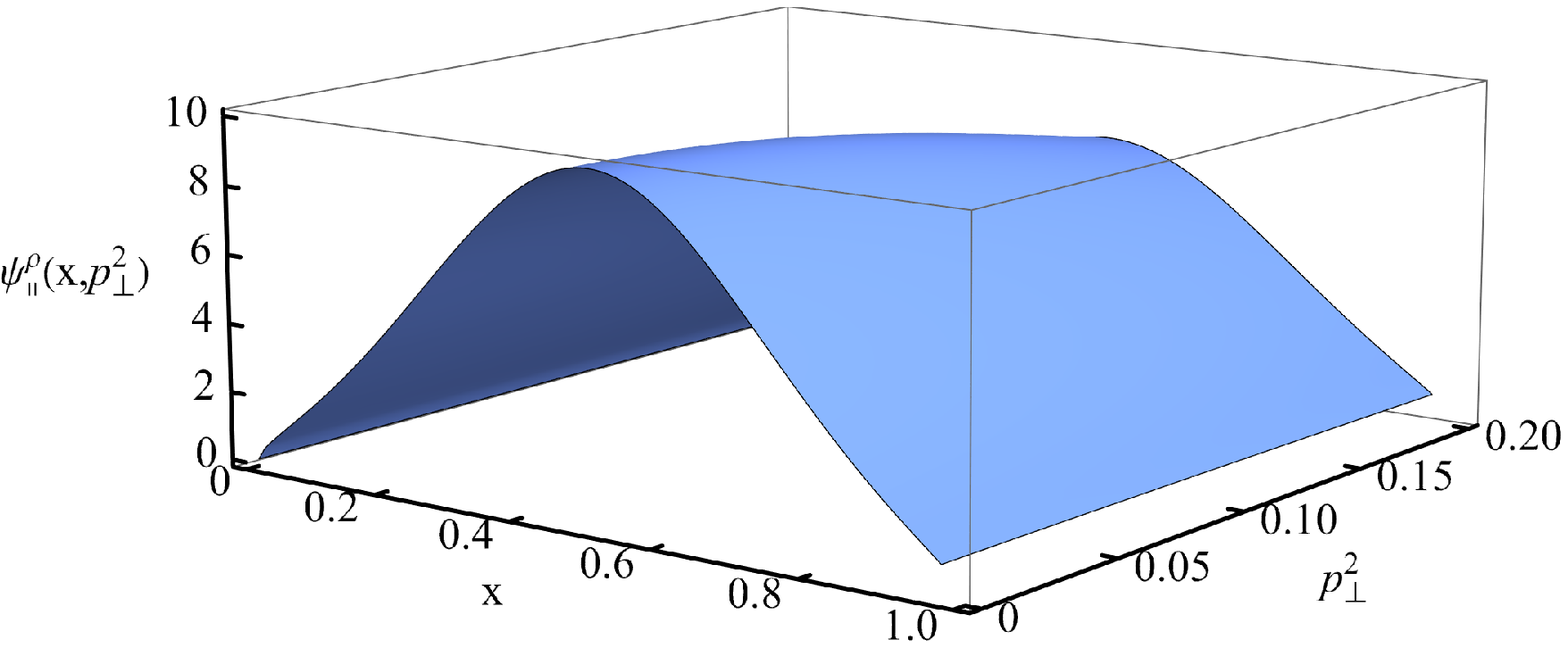}
\includegraphics[width=0.45\textwidth]{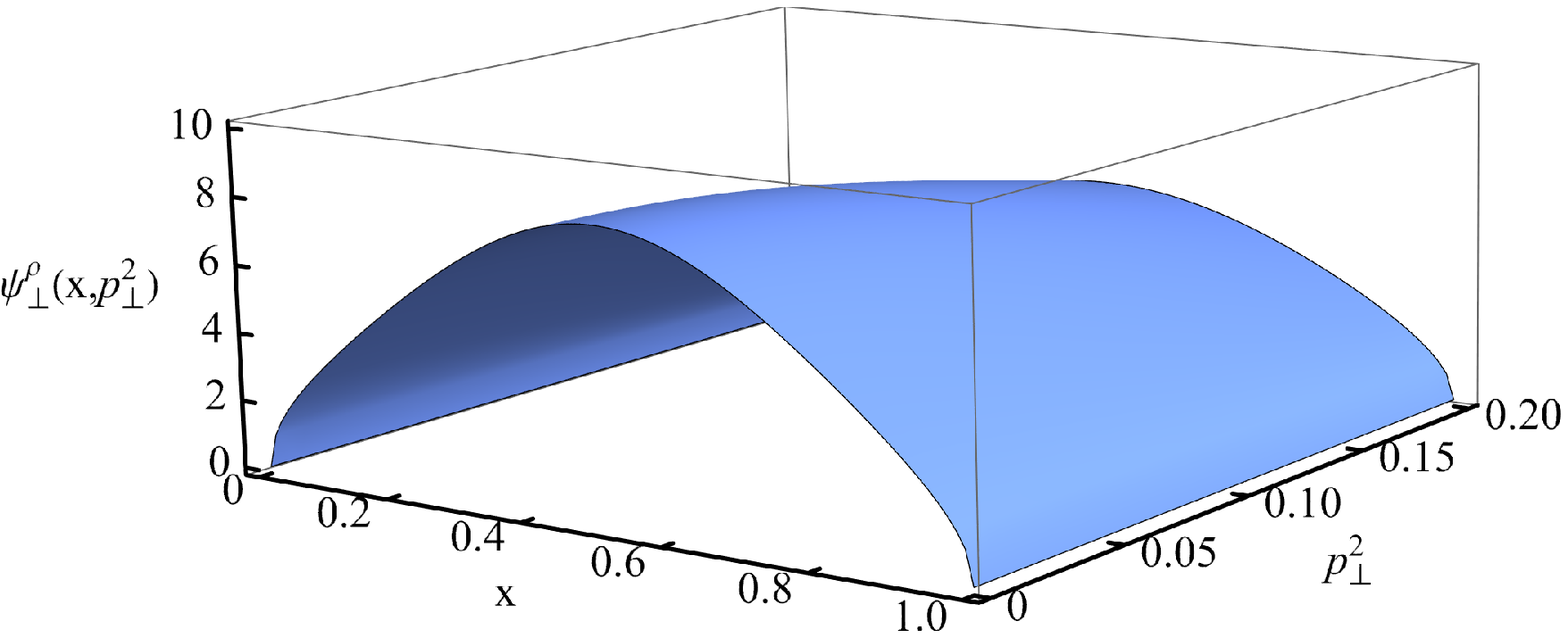} \\
\vspace*{0.60cm}
\includegraphics[width=0.45\textwidth]{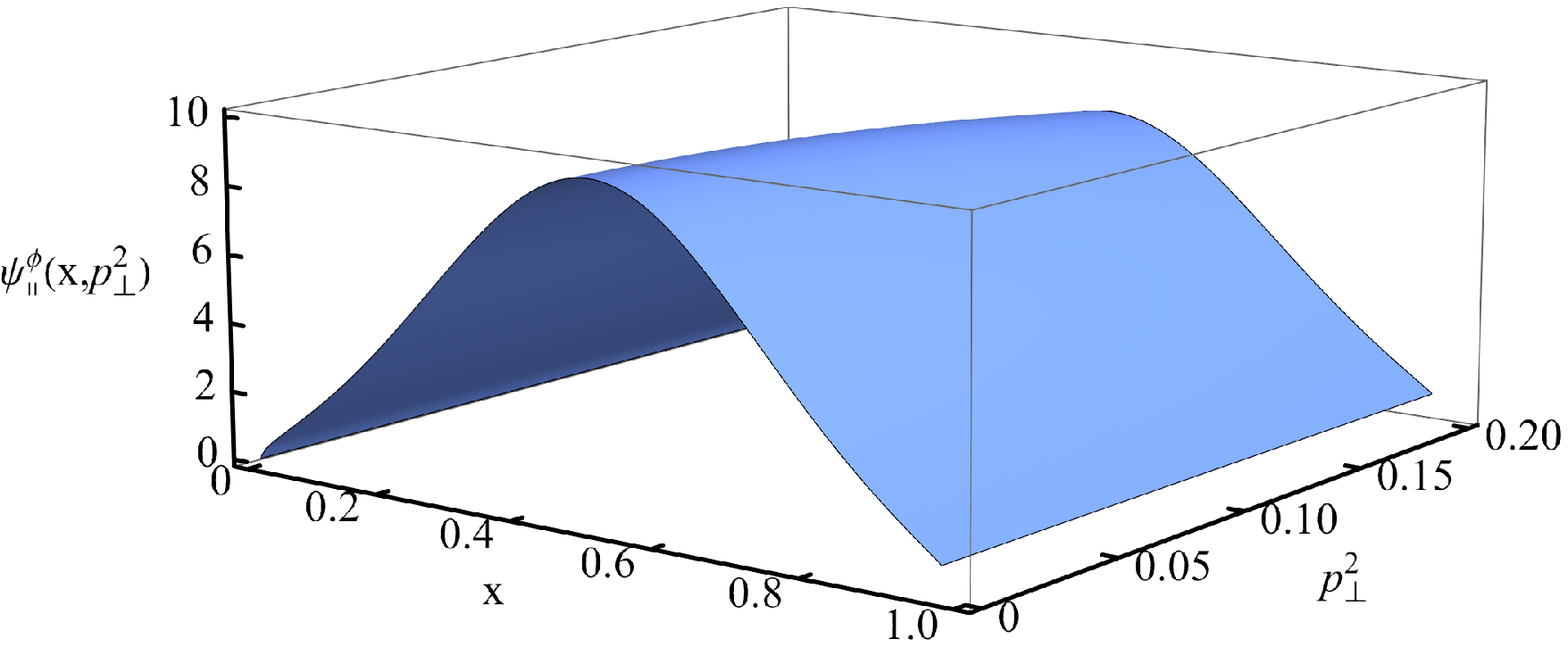}
\includegraphics[width=0.45\textwidth]{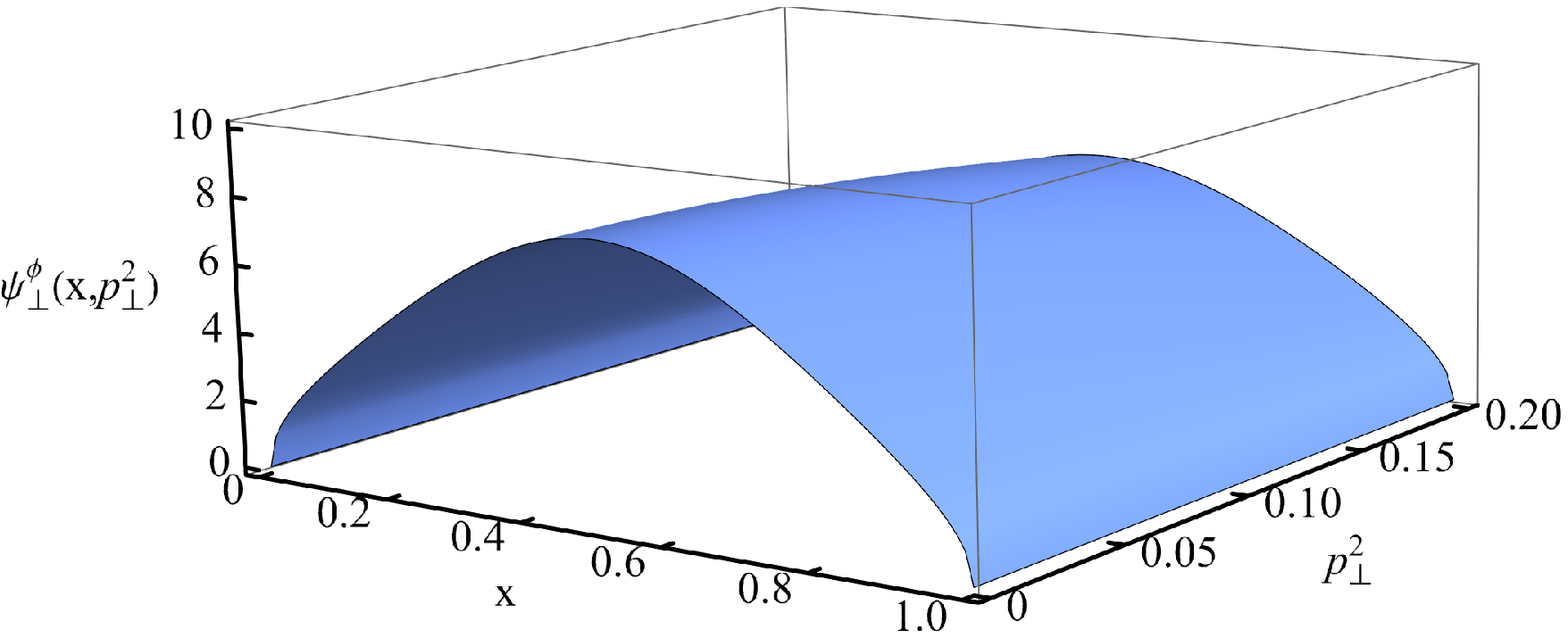}
\caption{\label{fig:LFWFs-Light} Longitudinal (left) and transverse (right) LFWFs of the $\rho$ (upper row) and $\phi$ (lower row) mesons.}
\end{figure*}

\begin{figure*}[t!]
\centering
\includegraphics[width=0.45\textwidth]{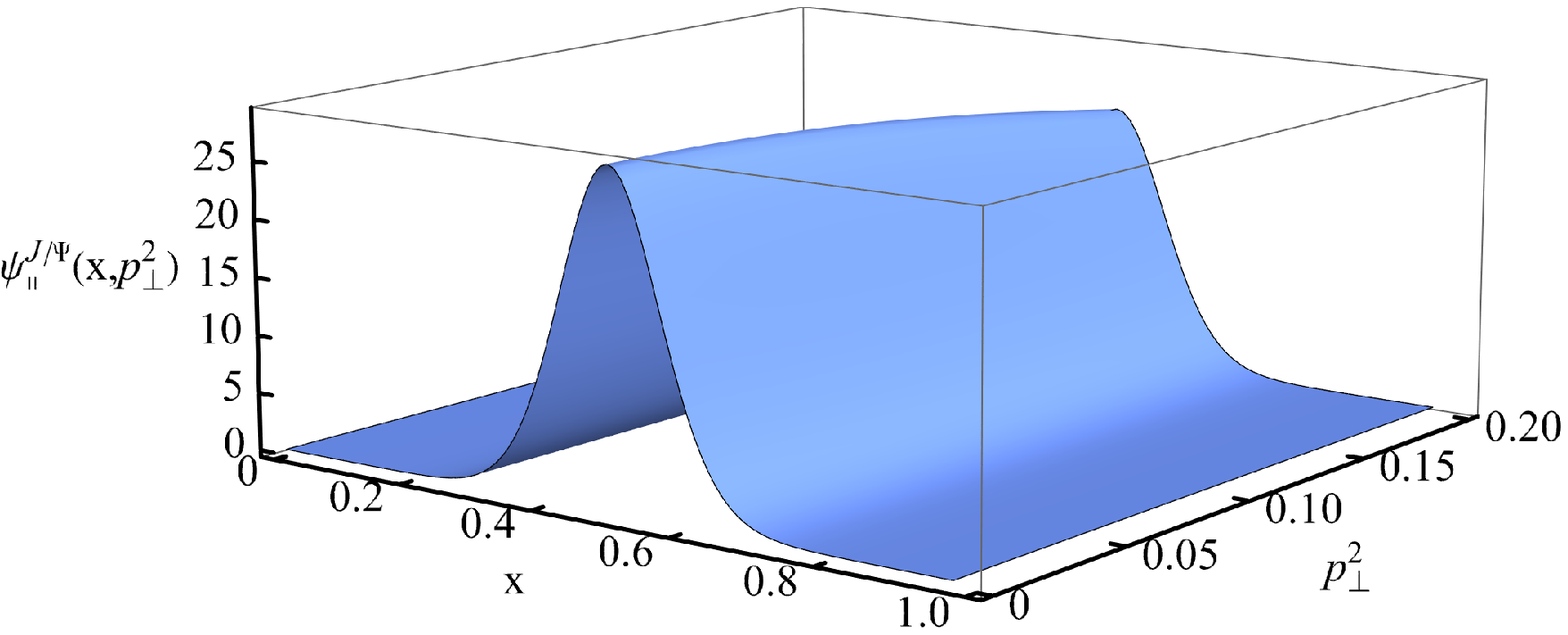}
\includegraphics[width=0.45\textwidth]{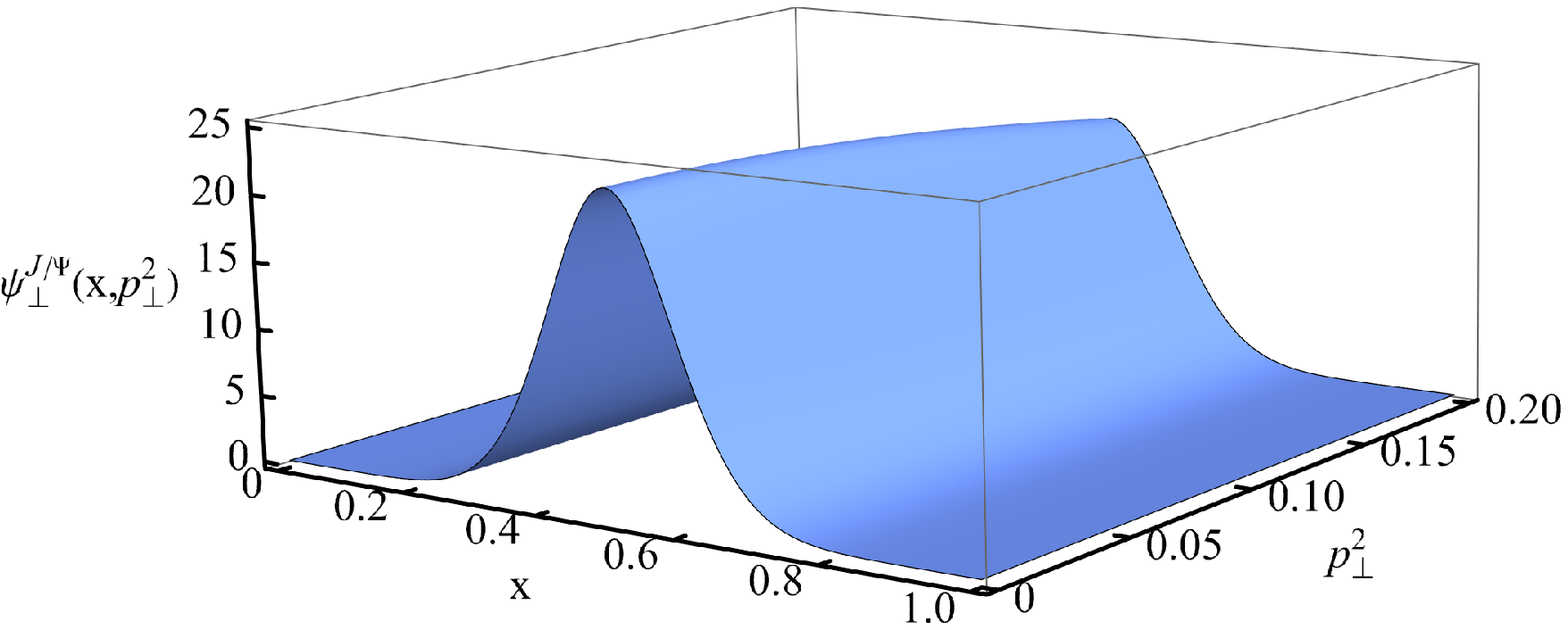} \\
\vspace*{0.60cm}
\includegraphics[width=0.45\textwidth]{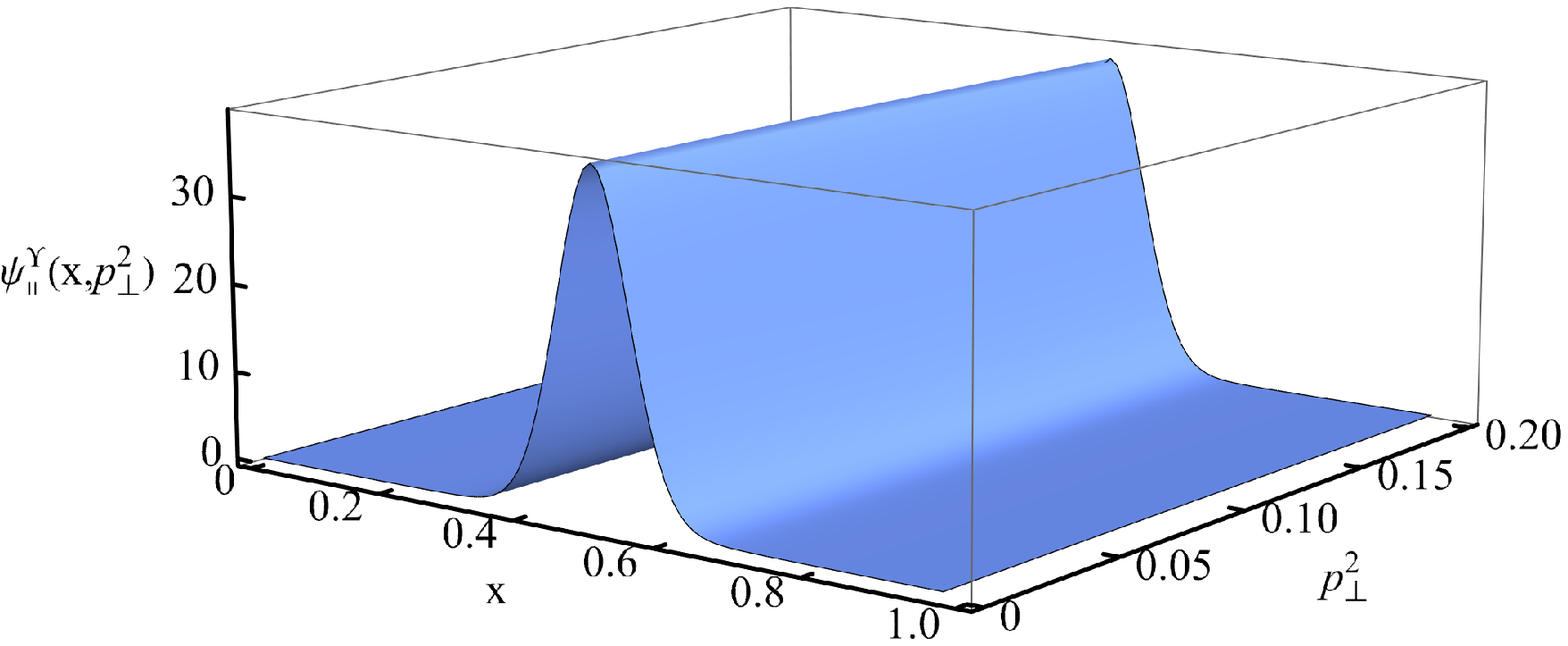}
\includegraphics[width=0.45\textwidth]{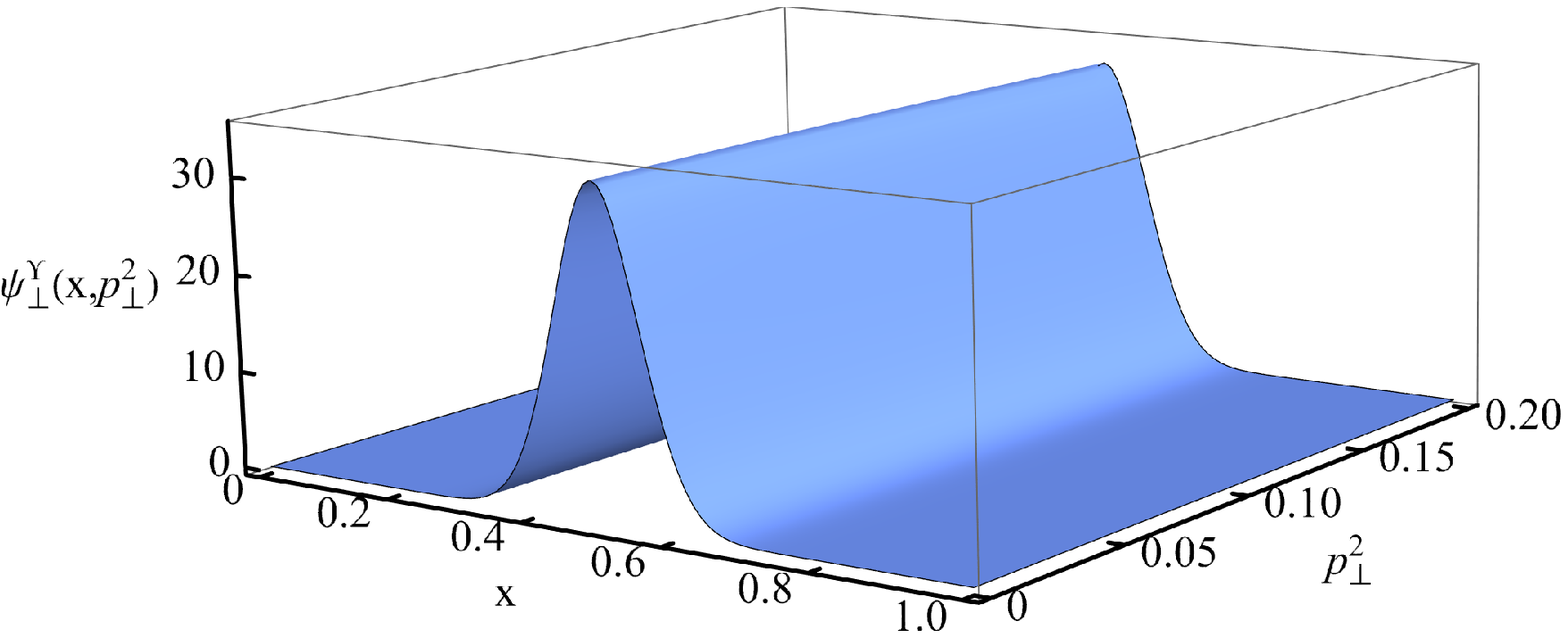}
\caption{\label{fig:LFWFs-Heavy} Longitudinal (left) and transverse (right) LFWFs of the $J/\psi$ (upper row) and $\Upsilon(1S)$ (lower row) mesons.}
\end{figure*}

Finally, Figs.~\ref{fig:LFWFs-Light} and~\ref{fig:LFWFs-Heavy} show the leading-twist longitudinal and transverse LFWFs of the lowest-lying vector mesons with hidden-light and -heavy flavor, respectively. One can see in Fig.~\ref{fig:LFWFs-Light} that the longitudinal and transverse LFWFs of the $\rho$ and $\phi$ mesons exhibit strong $x$- and $p_\perp^2$-dependence. The $\rho$-meson presents an $x$ point-wise behavior for the transverse component which is wider, and softer, than its longitudinal counterpart. Moreover, the fall-off along the $p_\perp^2$-range depicted is also smoother for a transversely polarized $\rho$-meson than for a longitudinal one. The $\phi$ meson exhibits LFWFs which are narrower in the $x$-dependence and flatter in the $p_\perp^2$-range depicted than those of the $\rho$ meson. This last feature is directly connected with the different mass of the dressed (anti-)quark within the $\rho$ and $\phi$ mesons. For both light vector mesons, it seems there is a difference with respect the $x$-dependence of the longitudinal and transverse LFWFs at their end points; while a slightly convex form can be deduced for the longitudinal part at the ends of the $x$-range, a concave trend is found for the transverse LFWF.

A different picture is shown in Fig.~\ref{fig:LFWFs-Heavy} for the lowest-lying vector mesons with hidden heavy flavor. Both longitudinal and transverse LFWFs are narrower, and harder, in $x$-dependence than those of light vector mesons. Furthermore, all LFWFs depicted are almost constant in the $p_\perp^2$-range depicted. With respect the $x$-dependence, it is worth noting that the longitudinal part seems to be even narrower than the transverse one in both $J/\psi$ and $\Upsilon(1S)$ mesons. This indicates that, centered at $x=1/2$, a wider range of meson's light-front momentum is available for the constituent heavy quark when such a meson is transversely polarized. It is also observed in Fig.~\ref{fig:LFWFs-Heavy} that the narrow feature of LFWFs is more prominent for the $\Upsilon(1S)$ meson than for the $J/\psi$ one, indicating that the features highlighted here are directly connected with the mass of the meson's constituents. Finally, in contrast with what was found in the light quark sector, it is clearly shown in Fig.~\ref{fig:LFWFs-Heavy} that the end points of both longitudinal and transverse LFWFs behave in approximately the same way with respect the $x$-dependence.

\begin{table*}[!t]
\caption{\label{tab:MM-LFWFs-0.0} Computed $\expval{(2x-1)^m}$ moments, with $m=0$, $2$, $\ldots$, $10$, of the leading-twist LFWFs at $p_\perp^2=0.0\,\text{GeV}^2$ for the vector mesons of hidden-flavor quark content. Odd moments are zero and all quantities are given in GeV$^{-2}$.}
\begin{ruledtabular}
\scalebox{0.90}{
\begin{tabular}{r|cccccc}
$\expval{(2x-1)^m}$ & $m=0$ & 2 & 4 & 6 & 8 & 10 \\
\hline
$\rho$ $\parallel$ & 5.48 & 0.92 & 0.37 & 0.20 & 0.13 & 0.09 \\
$\perp$            & 5.63 & 1.21 & 0.57 & 0.40 & 0.23 & 0.17 \\
\hline
$\phi$ $\parallel$ & 5.24 & 0.87 & 0.35 & 0.19 & 0.12 & 0.08 \\
$\perp$            & 5.32 & 1.14 & 0.53 & 0.32 & 0.22 & 0.16 \\ 
\hline
$J/\Psi$ $\parallel$ & 5.30 & 0.13 & 0.011 & 0.002 & $5.3\times 10^{-4}$ & $2.3\times 10^{-4}$ \\
$\perp$ & 5.30 & 0.19 & 0.021 & 0.004 & 0.001 & $3.7\times 10^{-4}$ \\
\hline
$\Upsilon(1S)$ $\parallel$ & 5.03 & 0.06 & 0.003 &$7.2\times 10^{-4}$ & $3.1\times 10^{-4}$ & $1.7\times 10^{-4}$ \\
$\perp$ & 5.03 & 0.07 & 0.003 & $2.5\times 10^{-4}$ & $2.6\times 10^{-5}$ & $3.6\times 10^{-6}$ \\
\end{tabular}
}
\end{ruledtabular}
\end{table*}

\begin{table*}[!t]
\caption{\label{tab:MM-LFWFs-0.1} Computed $\expval{(2x-1)^m}$ moments, with $m=0$, $2$, $\ldots$, $10$, of the leading-twist LFWFs at $p_\perp^2=0.1\,\text{GeV}^2$ for the vector mesons of hidden-flavor quark content. Odd moments are zero and all quantities are given in GeV$^{-2}$.}
\begin{ruledtabular}
\begin{tabular}{r|cccccc}
$\expval{(2x-1)^m}$ & $m=0$ & 2  & 4 & 6 & 8 & 10 \\ 
\hline
$\rho$ $\parallel$ & 5.10 & 0.87 & 0.35 & 0.19 & 0.12 & 0.08 \\
$\perp$            & 5.11 & 1.12 & 0.52 & 0.32 & 0.22 & 0.16 \\
\hline
$\phi$ $\parallel$ & 5.05 & 0.84 & 0.34 & 0.18 & 0.12 & 0.08 \\
$\perp$            & 5.06 & 1.09 & 0.51 & 0.31 & 0.21 & 0.16 \\
\hline
$J/\Psi$ $\parallel$ & 5.06 & 0.12 & 0.010 & 0.002 & $5.3\times 10^{-4}$ & $2.3\times 10^{-4}$ \\
$\perp$ & 5.06 & 0.18 & 0.020 & 0.004 & 0.001 & $3.7\times 10^{-4}$ \\
\hline
$\Upsilon(1S)$ $\parallel$ & 5.00 & 0.06 & 0.003 & $7.2\times 10^{-4}$ & $3.1\times 10^{-4}$ & $1.7\times 10^{-4}$ \\
$\perp$ & 5.00 & 0.07 & 0.003 & $2.4\times 10^{-4}$ & $2.6\times 10^{-5}$ & $3.6\times 10^{-6}$  \\
\end{tabular}
\end{ruledtabular}
\end{table*}

\begin{table*}[!t]
\caption{\label{tab:MM-LFWFs-0.2} Computed $\expval{(2x-1)^m}$ moments, with $m=0$, $2$, $\ldots$, $10$, of the leading-twist LFWFs at $p_\perp^2=0.2\,\text{GeV}^2$ for the vector mesons of hidden-flavor quark content. Odd moments are zero and all quantities are given in GeV$^{-2}$.}
\begin{ruledtabular}
\begin{tabular}{r|cccccc}
$\expval{(2x-1)^m}$ & $m=0$ & 2 & 4 & 6 & 8 & 10 \\
\hline
$\rho$ $\parallel$ & 4.22 & 0.74 & 0.30 & 0.17 & 0.11 & 0.07 \\
$\perp$            & 3.92 & 0.89 & 0.42 & 0.26 & 0.18 & 0.13 \\
\hline
$\phi$ $\parallel$ & 4.55 & 0.77 & 0.31 & 0.17 & 0.11 & 0.08 \\
$\perp$            & 4.41 & 0.97 & 0.46 & 0.28 & 0.19 & 0.14 \\
\hline
$J/\Psi$ $\parallel$ & 4.45 & 0.11 & 0.009 & 0.002 & $5.1\times 10^{-4}$ & $2.2\times 10^{-4}$ \\
$\perp$ & 4.44 & 0.16 & 0.190 & 0.004 & $9.9\times 10^{-4}$ & $3.5\times 10^{-4}$\\
\hline
$\Upsilon(1S)$ $\parallel$ & 4.90 & 0.06 & 0.003 & $7.1\times 10^{-4}$ &$3.1\times 10^{-4}$ & $1.7\times 10^{-4}$ \\
$\perp$ & 4.90 & 0.07 & 0.003 & $2.4\times 10^{-4}$ & $2.6\times 10^{-5}$ & $3.6\times 10^{-6}$ \\
\end{tabular}
\end{ruledtabular}
\end{table*}

For completeness, Tables~\ref{tab:MM-LFWFs-0.0},~\ref{tab:MM-LFWFs-0.1} and~\ref{tab:MM-LFWFs-0.2} show the computed $\expval{(2x-1)^m}$ moments, with $m=0$, $2$, $\ldots$, $10$, of the leading-twist LFWFs at $p_\perp^2=0.0$, $0.1$ and $0.2$, respectively, for the vector mesons of hidden-flavor quark content, \emph{i.e.}, the $\rho(770)$, $\phi(1020)$, $J/\psi$ and $\Upsilon(1S)$ mesons. The most salient features of this analysis are: (i) moments of heavy mesons are already very small from $m=2-4$ whereas one must go beyond the $m=10$ moments in order to obtain a similar order of magnitude for light vector mesons; (ii) the value of a given moment generally decreases as $p_\perp^2$ increases for any vector meson, once the moment's value is small enough it remains nearly constant with respect to changes in $p_\perp^2$; and (iii) differences between the values of moments are found for the longitudinal and transverse LFWFs, such differences are more pronounced if either the order of the moment is higher or the vector meson is lighter.


\section{Summary}
\label{sec:summary}

Following a recently proposed algebraic model which adequately describes the internal structure of the lowest-lying hidden-flavor pseudo-scalar mesons, with either light or heavy quark content, we have performed its first extension to the vector-meson case, calculating the leading-twist LFWFs of the $\rho(770)$, $\phi(1020)$, $J/\psi$ and $\Upsilon(1S)$ mesons from their respective valence-quark (twist-two) PDAs. 

The PDA is usually easy to extract within a given theoretical framework due to its simplicity. However, this is not the case for the LFWF which is the closest physical object in a quantum field theory to the notion of a wave function in a non-relativistic system. Moreover, the comparison of the results reported herein with those of pseudo-scalar mesons highlights features of the mechanism that drives DCSB within these two systems.

The connection between vector meson LFWFs and PDAs has been derived, computed numerically, plotted and analyzed in great detail in this work, concluding that the transverse LFWFs can be obtained directly from the corresponding PDA, whereas the calculation of the longitudinal LFWF requires the knowledge of the SDF derived from its corresponding PDA. This last feature is not necessary a drawback of our theoretical framework; for instance, the SDF, and its modification, can also serve as a benchmark for investigating the transition between meson's vacuum properties and the change they undergo when dealing with finite temperature and density.

With respect to our results on the SDFs, the $\rho$ and $\phi$ mesons present very similar SDFs, both are positive definite in the whole range, with non-zero end points and a point-wise behavior that alternates convex and concave forms which resemble the ones already found for the lowest-lying pseudo-scalar mesons. As far as the SDFs of the $J/\psi$ and $\Upsilon(1S)$ are concerned, they are narrower than those of light vector mesons and, furthermore, they are not positive definite along the entire interval, although the area under the curve is larger than zero.

Concerning the LFWFs studied herein, the $\rho$-meson exhibits an $x$-dependence for the transverse component which is wider, and softer, than its longitudinal counterpart. Moreover, the fall-off along the $p_\perp^2$-range depicted is smoother for a transversely polarized $\rho$-meson than for a longitudinal one. The $\phi$ meson exhibits LFWFs which are narrower in the $x$-dependence and flatter in the $p_\perp^2$-range than those of the $\rho$-meson. The longitudinal and transverse LFWFs of the lowest-lying vector mesons with hidden heavy flavor are observed to be steeper in the $x$-dependence than those of light vector mesons; besides, the longitudinal part seems to be even narrower than the transverse one in both $J/\psi$ and $\Upsilon(1S)$ mesons. Furthermore, all LFWFs of heavy vector mesons are nearly constant in the $p_\perp^2$-range depicted.


\begin{acknowledgments}
%
B. Almeida Zamora acknowledges CONACyT (No. CVU 935777) for PhD fellowship.
J.J. Cobos-Martínez acknowledges financial support from the University of Sonora under grant USO315007861.
%
A. Bashir acknowledges Coordinaci\'on de la Investigaci\'on Cient\'ifica of the Universidad Michoacana de San Nicol\'as de Hidalgo Grant No. 4.10., the US Department of Energy (DOE) under the Contract No. DE-AC05-6OR23177 and the Fulbright-Garc\'ia Robles Scholarship.
This work has also been partially funded by Ministerio Espa\~nol de Ciencia e Innovaci\'on under grant No. PID2019-107844GB-C22; Junta de Andaluc\'ia under contract Nos. Operativo FEDER Andaluc\'ia 2014-2020 UHU-1264517, P18-FR-5057 and also PAIDI FQM-370.
\end{acknowledgments}


\appendix

\begin{figure*}[!t]
\begin{center}
\includegraphics[width=0.45\textwidth]{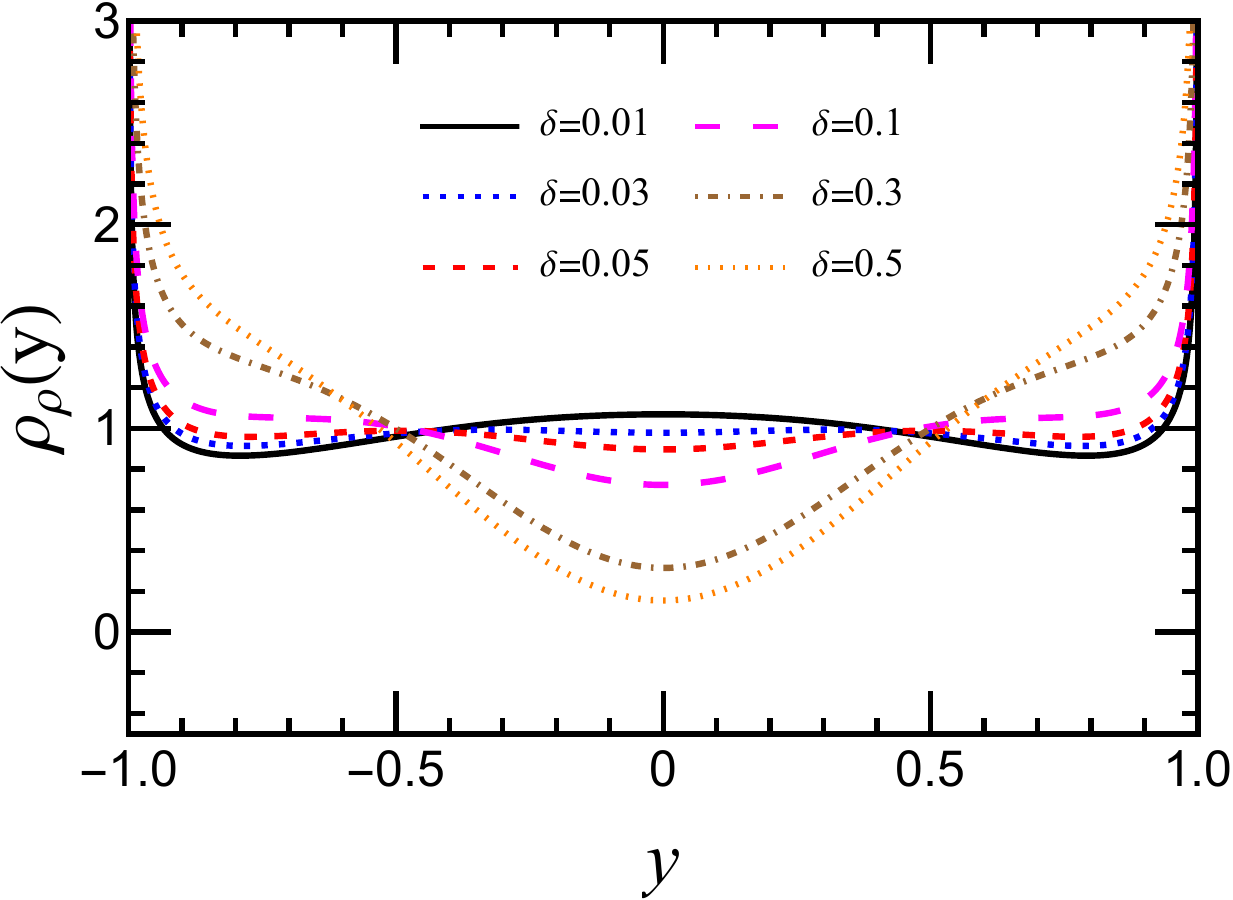}
\includegraphics[width=0.47\textwidth]{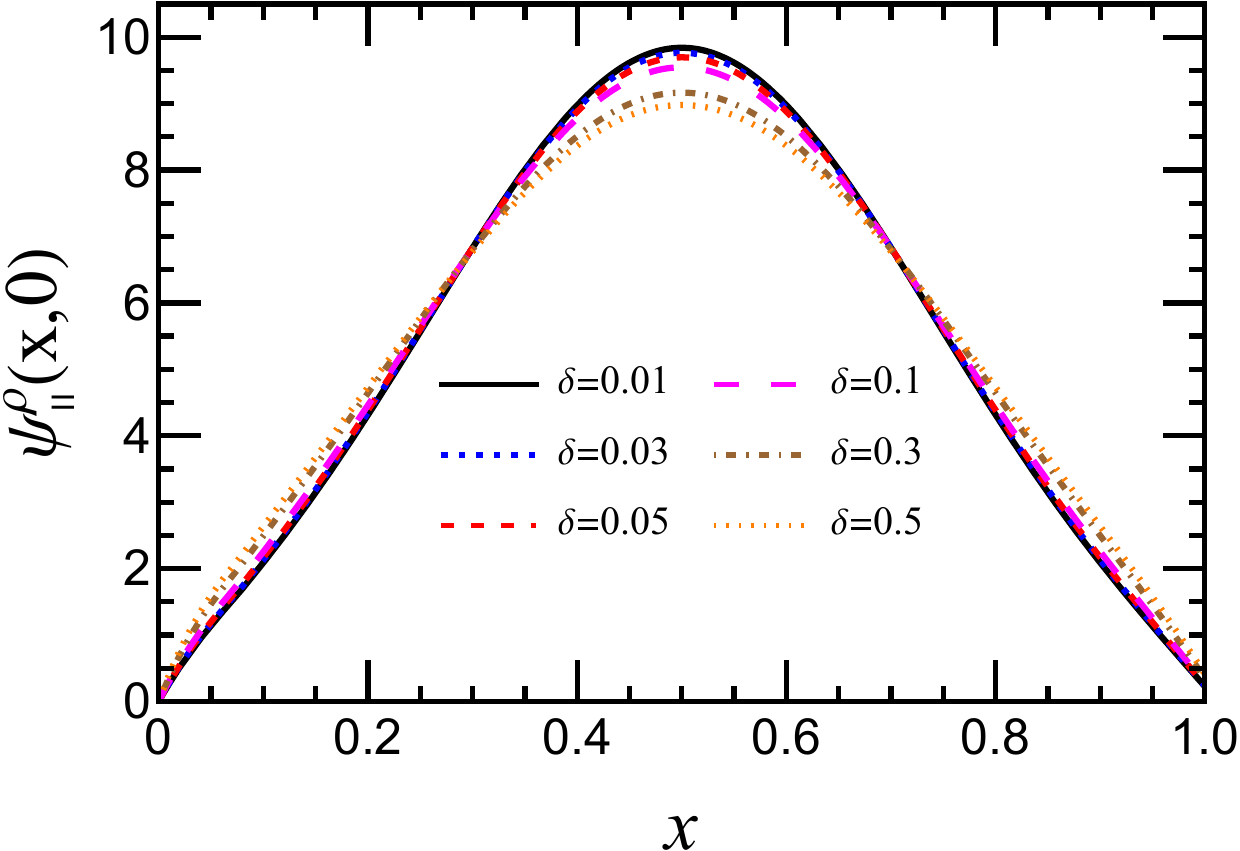} \\
\includegraphics[width=0.47\textwidth]{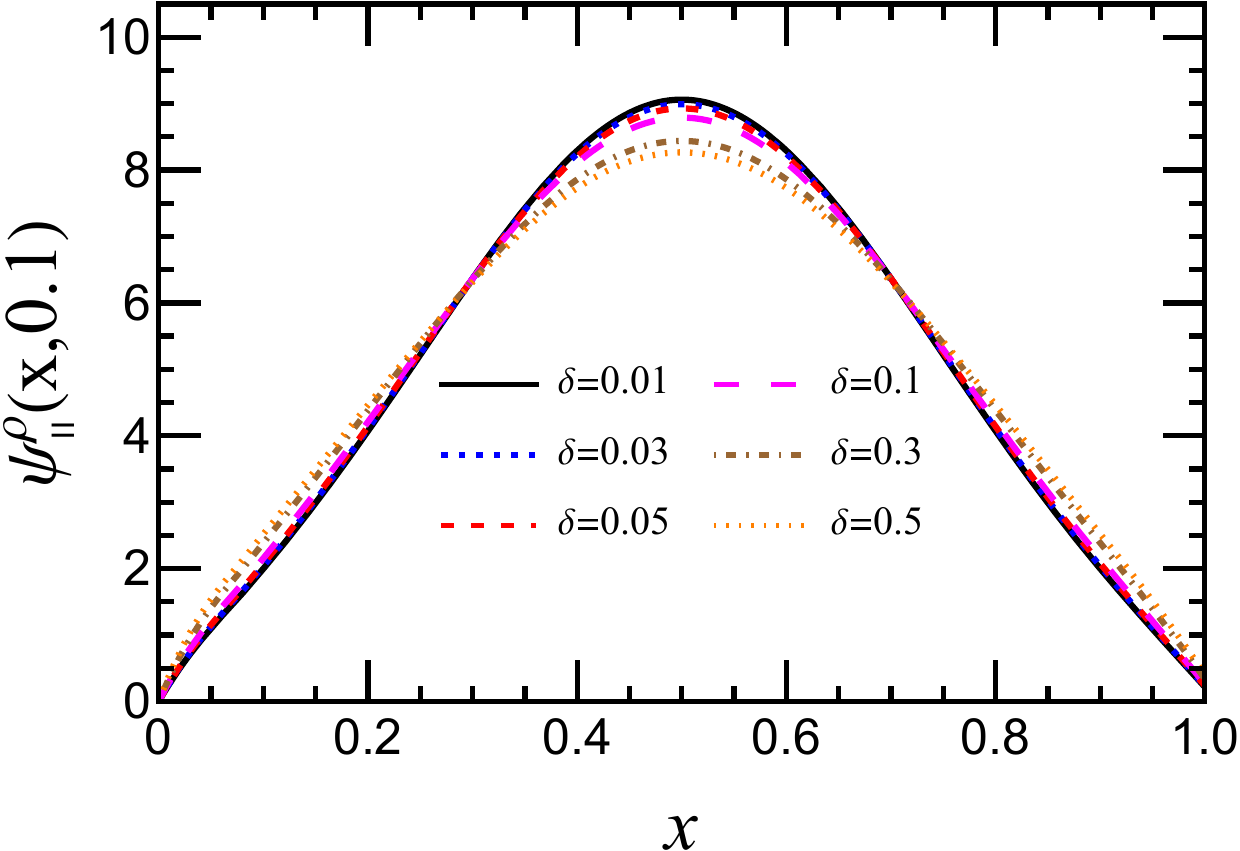}
\includegraphics[width=0.47\textwidth]{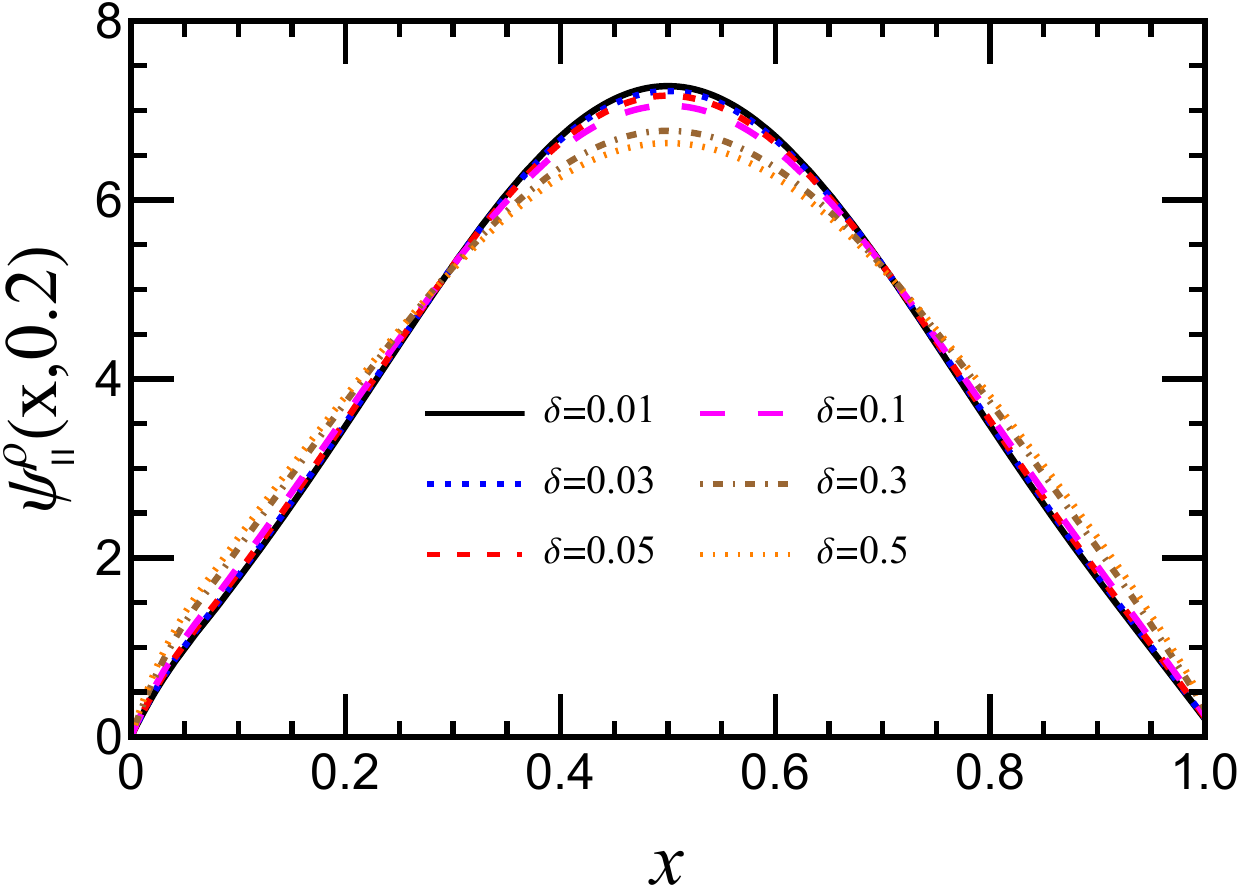}
\end{center}
\vspace*{-0.40cm}
\caption{\label{fig:rho-deltas} Spectral density and light-front wave functions of $\rho$-meson calculated for different values of $\delta=\nu-1$. The LFWF is plotted for $p_\perp=0.0$, $0.1$ and $0.2$ in order to analyze in detail differences related with the $p_\perp$-dependence.}
\end{figure*}

\begin{figure*}[!t]
\begin{center}
\includegraphics[width=0.48\textwidth]{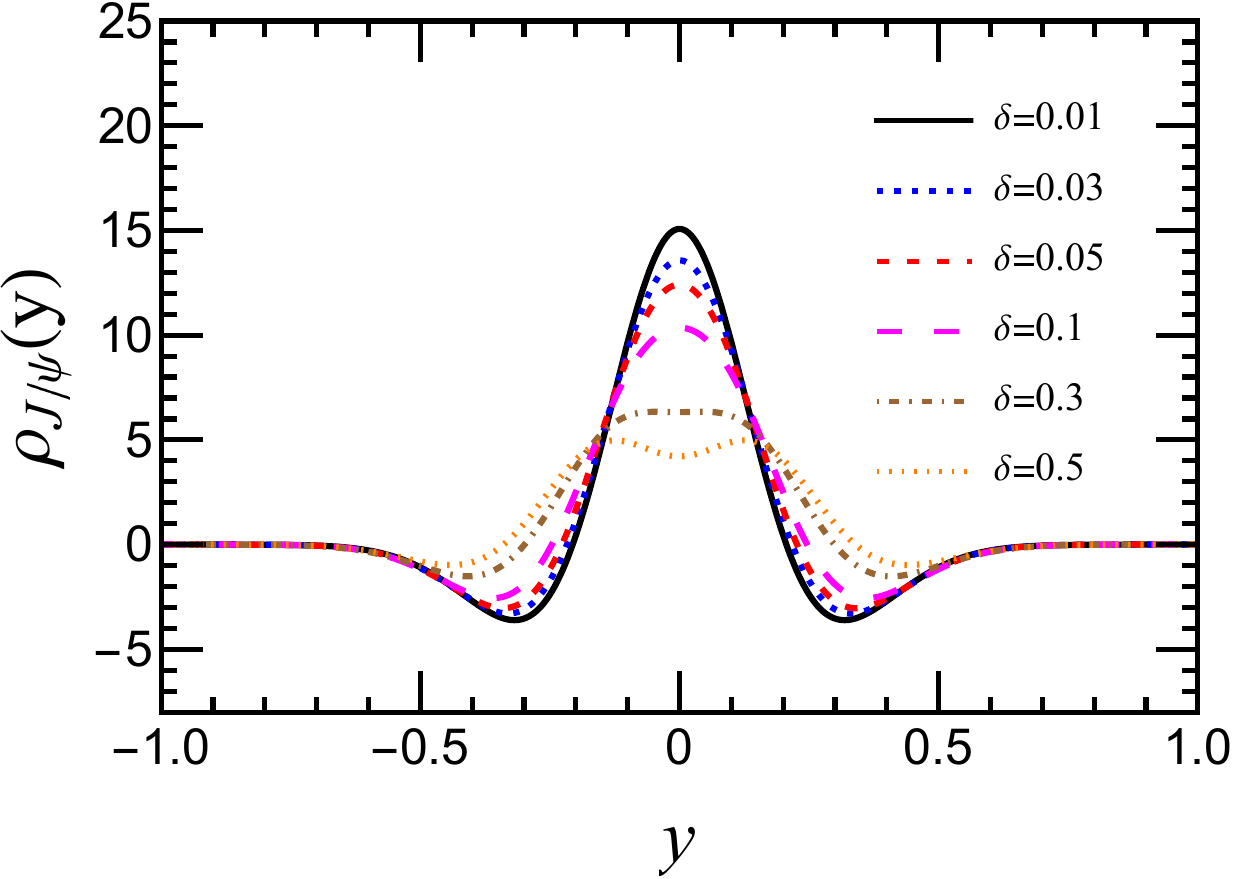}
\includegraphics[width=0.47\textwidth]{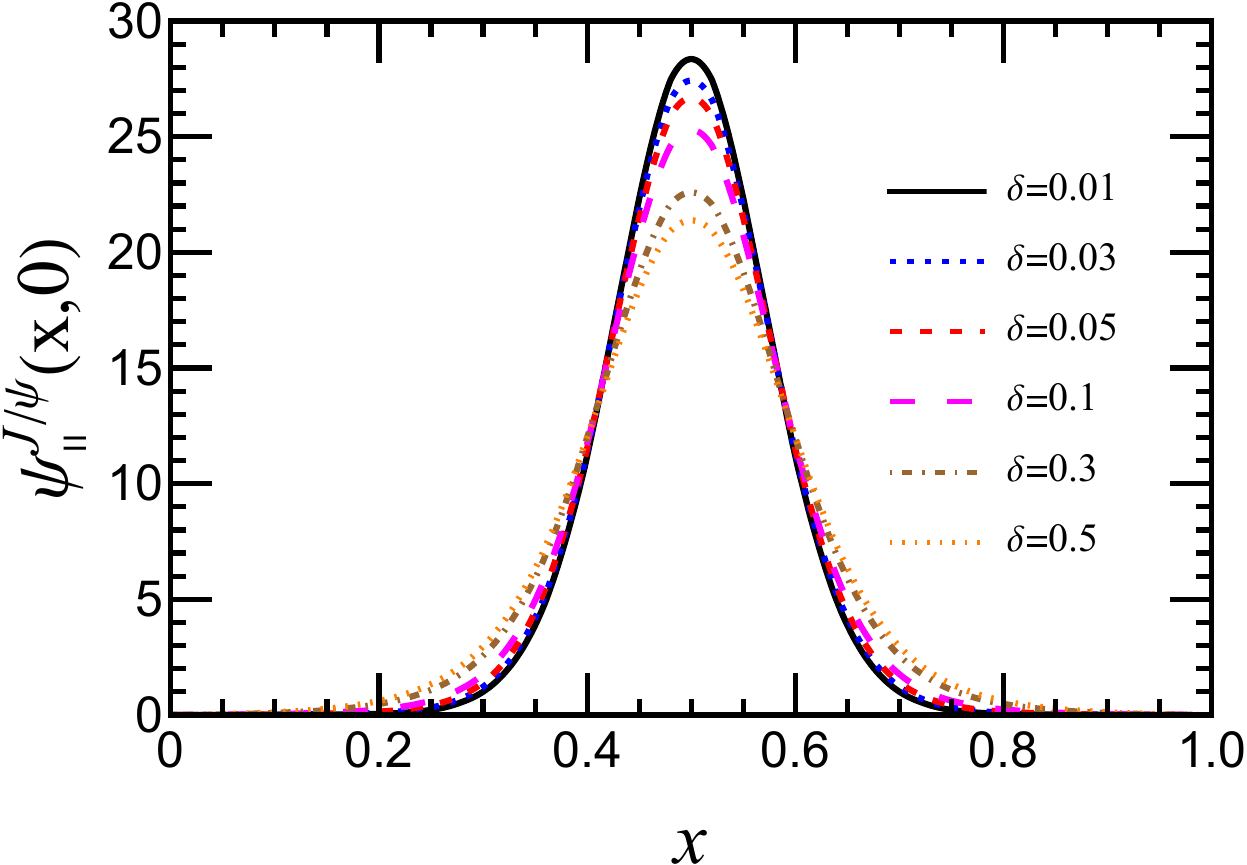} \\
\includegraphics[width=0.47\textwidth]{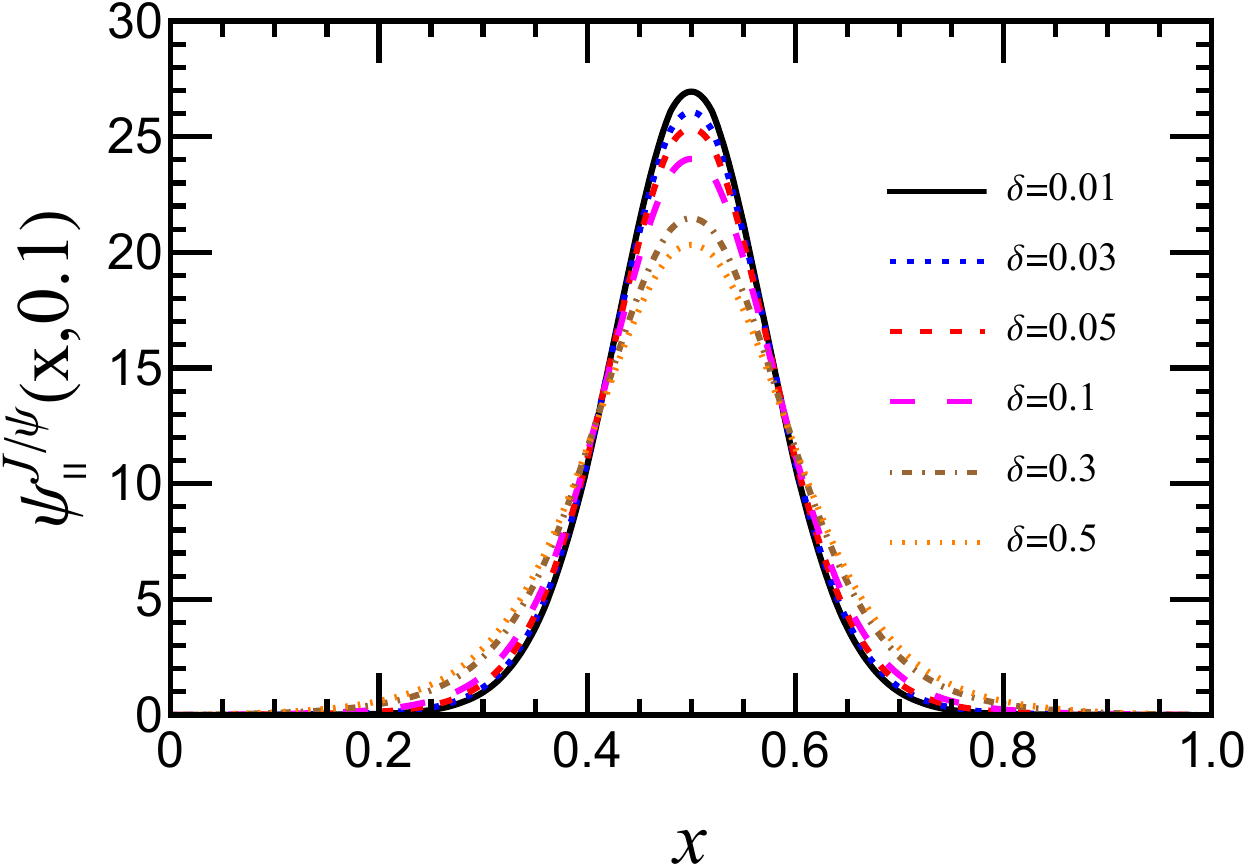}
\includegraphics[width=0.47\textwidth]{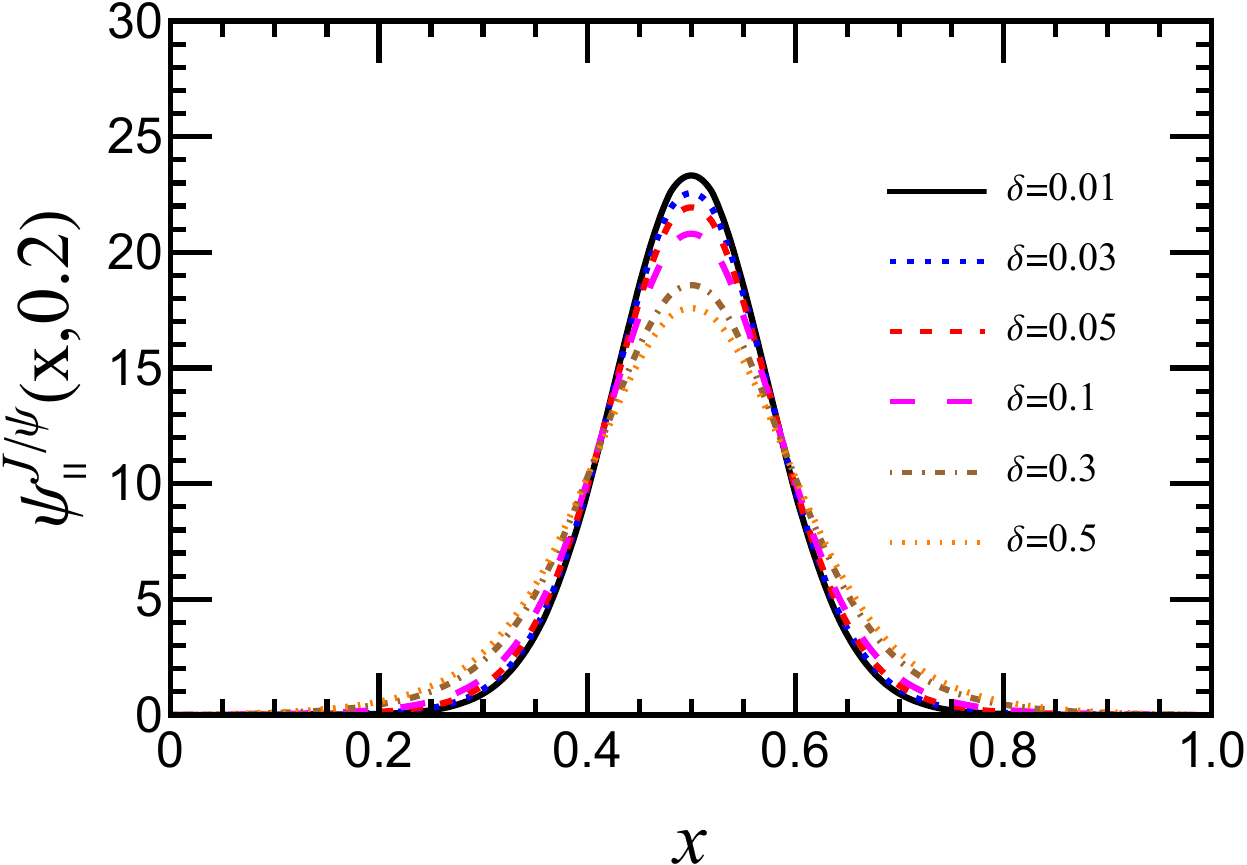}
\end{center}
\vspace*{-0.40cm}
\caption{\label{fig:jpsi-deltas} Spectral density and light-front wave functions of $J/\psi$-meson calculated for different values of $\delta=\nu-1$. The LFWF is plotted for $p_\perp=0.0$, $0.1$ and $0.2$ in order to analyze in detail differences related with the $p_\perp$-dependence.}
\end{figure*}

\section{On the choice of $\mathbf{\nu}$ and LFWFs}

In order to analyze how sensitive our results are to the particular choice of $\delta=0.01$, we change it by the following ones: $0.03$, $0.05$, $0.1$, $0.3$ and $0.5$; that is to say, we enlarge it by up to $50$ times its original value. This modification affects mostly the point-wise behavior of the SDF but it has little effect on the $x$- and $p_\perp$-dependence of the LFWF. This can be seen in Figs.~\ref{fig:rho-deltas} and~\ref{fig:jpsi-deltas} for the $\rho(770)$ and $J/\psi$ mesons, and similar results are obtained for the corresponding heavier ones, \emph{i.e.} the $\phi(1020)$ and $\Upsilon(1S)$ mesons.

The upper-left panel of Fig.~\ref{fig:rho-deltas} shows that the SDF is sensitive to the variation of the $\delta$-parameter. The drastic change in SDF is due to the fact that we have multiplied the $\delta$'s value by a factor of $50$. However, as shown in the remaining panels of Fig.~\ref{fig:rho-deltas}, such a large modification of the $\delta$-parameter produces little effect in our prediction of the point-wise behavior of the longitudinal LFWF. One may conclude that the change in the peak is around $10\%$, and even less on the ends, when modifying heavily the $\delta$-parameter. The case of vector mesons with heavy-quark-antiquark content is drawn in Fig.~\ref{fig:jpsi-deltas}, one can observe that, despite the SDF behaves similarly as in the light vector meson case, the longitudinal LFWF is more sensitive, with a maximum change in the peak of around $30\%$, and even less on their sides. We consider this variation tolerable for inflating the $\delta$-parameter $50$ times. 

All this allows us to conclude that our prediction for the LFWFs of the $\rho(770)$, $\phi(1020)$, $J/\psi$ and $\Upsilon(1S)$ mesons are reasonably robust within our theoretical approach, and assumptions.


\bibliography{PrintLFWFsVectorMesons}

\end{document}